\documentclass[nofootinbib, amsmath, amssymb, 10pt, eqsecnum, twocolumn, a4paper]{revtex4-1}
\usepackage{amsfonts}
\usepackage{amssymb}
\usepackage{amsmath}
\usepackage[hmargin=2cm,vmargin=2cm]{geometry}
\usepackage{color}
\usepackage{graphicx}
\usepackage{rotating}
\usepackage[breaklinks, colorlinks, citecolor=blue]{hyperref}
\usepackage[greek,english]{babel}
\usepackage{enumitem}
\usepackage{multirow}
\usepackage{natbib,hyperref,ifthen}
\usepackage[toc,page]{appendix}
\usepackage{array}

\newcolumntype{P}[1]{>{\centering\arraybackslash}p{#1}}

\usepackage{nameref}
\usepackage{chngcntr}
\counterwithout{paragraph}{subsubsection}
\counterwithin*{paragraph}{subsection}

\newcommand{\btheta}{{\boldsymbol\theta}}
\newcommand{\dtheta}{\boldsymbol{\delta\theta}}
\newcommand{\CMB}{\rm CMB}
\newcommand{\LSS}{\rm LSS}

\begin{document}
\author{Tom Charnock}
\email{tom.charnock@nottingham.ac.uk}
\affiliation{Centre for Astronomy \& Particle Theory, University of Nottingham, University Park, Nottingham, NG7 2RD, U.K.}
\author{Richard A. Battye}
\email{richard.battye@manchester.ac.uk}
\affiliation{Jodrell Bank Centre for Astrophysics, School of Physics and Astronomy, University of Manchester, Manchester, M13 9PL, U.K.}
\author{Adam Moss}
\email{adam.moss@nottingham.ac.uk}
\affiliation{Centre for Astronomy~\& Particle Theory, University of Nottingham, University Park, Nottingham, NG7 2RD, U.K.}
\title{\emph{Planck} confronts large scale structure:\\methods to quantify discordance}

\begin{abstract}
	\noindent Discordance in the {\greektext L}CDM cosmological model can be seen by comparing parameters constrained by CMB measurements to those inferred by probes of large scale structure. 
	Recent improvements in observations, including final data releases from both \emph{Planck} and SDSS-III BOSS, as well as improved astrophysical uncertainty analysis of CFHTLenS, allows for an update in the quantification of any tension between large and small scales.
	This paper is intended, primarily, as a discussion on the quantifications of discordance when comparing the parameter constraints of a model when given two different data sets.
	We consider KL-divergence, comparison of Bayesian evidences and other statistics which are sensitive to the mean, variance and shape of the distributions.
	However, as a by-product, we present an update to the similar analysis in \cite{Battye:2014qga} where we find that, considering new data and treatment of priors, the constraints from the CMB and from a combination of LSS probes are in greater agreement and any tension only persists to a minor degree. 
	In particular, we find the parameter constraints from the combination of LSS probes which are most discrepant with the \emph{Planck}2015+Pol+BAO parameter distributions can be quantified at a $\sim2.55\sigma$ tension using the method introduced in \cite{Battye:2014qga}.
	If instead we use the distributions constrained by the combination of LSS probes which are in greatest agreement with those from \emph{Planck}2015+Pol+BAO this tension is only $0.76\sigma$\footnote{The code used to calculate results in this paper can be found at \url{https://github.com/tomcharnock/tension}}.
\end{abstract}
\maketitle

	\section{Introduction}

	{\greektext L}CDM is an extremely successful cosmological model based on general relativity with components of dark energy and cold dark matter, in addition to baryonic matter.
	It predicts baryon acoustic oscillations (BAO), the periodic fluctuations in the density of visible matter, as well as the polarisation and gravitational lensing of photons.
	The evolution of {\greektext L}CDM is imprinted in the cosmic microwave background (CMB) radiation, relic light from the epoch of recombination, and in the large scale structure (LSS) of the universe and can be measured by a host of different techniques.
	{\greektext L}CDM can be quantified by just six cosmological parameters~\cite{Ade:2015xua}: the physical densities of baryonic and cold dark matter $\Omega_{\rm b}h^2$ and $\Omega_{\rm c}h^2$, the angular diameter of the CMB acoustic scale $\Theta_{\rm MC}$, the amplitude of the curvature density fluctuations $A_{\rm s}$, the scalar spectral index $n_{\rm s}$, and the optical depth to reionisation $\tau$.
	\\\\
	Measurements from the CMB or LSS can be used to constrain the {\greektext L}CDM parameters from which all other derived parameters can be calculated, e.g. the Hubble parameter, the physical density of matter or the amplitude of density fluctuation at a scale of 8$h^{-1}$Mpc, $H_0$, $\Omega_{\rm m}$ and $\sigma_8$ respectively.
	Indications of a discrepancy between constraints of the {\greektext L}CDM parameters when using either the CMB \emph{or} LSS probes could show that physics may deviate from \emph{vanilla} {\greektext L}CDM, i.e. with no additional physics.
	It has been widely noted~\cite{2013arXiv1303.5080P,2013arXiv1306.4153R,Battye:2013,Gao:2013pfa,Hamann:2013,Wyman:2013,Beutler:2014yhv,MacCrann:2014wfa,Ruiz:2014hma,Addison:2015wyg,Dossett:2015nda,Hu:2015rva,Kitching:2016hvn} that probes of LSS suggest the joint values of $\Omega_{\rm m}$ and $\sigma_8$ do not seem to agree with those obtained using the CMB.
	In particular, the constraints from LSS imply that there is too much small scale structure when compared to the constraints from measurements of the CMB.
	\\\\
	In \cite{Battye:2014qga} we analysed the discrepancy which arised in each of the five relevant {\greektext L}CDM parameters when using measurements of the CMB and observations of LSS available at that time.
	The CMB measurements were obtained using the \emph{Planck}2013 temperature anisotropies~\cite{Ade:2013zuv} combined with WMAP polarisation (WP)~\cite{Hinshaw:2012aka} and BAO results from the third Sloan Digital Sky Survey (SDSS-III) experiment, Baryon Oscillation Spetroscopic Survey (BOSS) DR9~\cite{Anderson:2012sa}, as well as the 6dF Galaxy Survey~\cite{Jones:2004zy} and the SDSS DR7 Main Galaxy Sample~\cite{Padmanabhan:2012hf}.
	A host of large scale structure probes were used in \cite{Battye:2014qga}, such as weak lensing from the Canada-France-Hawaii Telescope Lensing Survey (CFHTLenS)~\cite{Heymans:2013fya}, lensing of the CMB from the \emph{Planck} lensing analysis~\cite{Ade:2013tyw}, redshift-space distortions (RSD) from SDSS-III BOSS DR11~\cite{Beutler:2013yhm} and Sunyaev-Zel'dovich (SZ) galaxy cluster counts~\cite{2013arXiv1303.5080P}.
	By placing Gaussian priors from \emph{Planck}2013 on $\Theta_{\rm MC}$ and $n_{\rm s}$ in the Markov-chain Monte Carlo (MCMC) analysis from {\ttfamily COSMOMC}~\cite{Lewis:2002ah}, constraints using a combination of all the LSS probes could be obtained.
	By comparing the joint probability distribution for the five {\greektext L}CDM parameters from both the CMB and LSS we were able to state that this tension was in excess of $5\sigma$.
	This lead to the consideration of extensions to {\greektext L}CDM to reduce the power of density fluctuations on small scales only.
	This included adding either an active or sterile neutrino component, or modifying the tilt of the primordial power spectrum.
	By including these modifications to the model, the tension between the parameters obtained from the CMB and from LSS decreased to $\sim2\sigma$ reducing the discrepancy by a large amount, although not alleviating it totally.
	This came with a reasonably significant non-zero mass of active or sterile neutrinos, which was discrepant with the result obtained by \emph{Planck}~\cite{Ade:2013zuv}.
	\\\\
	This paper is primarily concerned with studying various methods to quantify any discrepancy which arises when two datasets predict different parameter values from the same model.
	In Sec.~\ref{S:Tension} we review (not exhaustively) different measures that can be used to interpret these discordances.
	We also define two methods which are not used in the literature which are robust in a wide variety of situations.
	A detailed explanation of how each of these measures perform when applied to a range of different probability distributions is presented in Appendix~\ref{Appendix}.
	In Sec.~\ref{S:Data} we will introduce the data used to constrain the {\greektext L}CDM parameters.
	Finally, in Sec.~\ref{S:Discordance} the remaining discordance between {\greektext L}CDM parameters when measured using the CMB and a combination of LSS probes is updated.
	We quote the amount of disagreement using the measures introduced in Sec.~\ref{S:Tension} and comment on how each of them can be interpreted.
	Finally in Sec.~\ref{Disc}, extending {\greektext L}CDM with neutrino content is discussed, as is a note on how the impact of the most recent \emph{Planck}2016\footnote{The MCMC chains or likelihood analysis for \emph{Planck}2016 was not available at the time of submission.} analysis may affect the results.
		
	\section{Quantifying Discrepancy\label{S:Tension}}

    The probability distribution $P(\btheta)$ of the five relevant {\greektext L}CDM parameters, excluding $\tau$ which is only constrained by the CMB, is a complicated 5D, not-necessarily Gaussian, function.
    When constraining the parameters using the CMB only, one distribution $P(\btheta|\CMB,\Lambda{\rm CDM})$ is found and a second, supposedly similar, distribution $P(\btheta|\LSS,\Lambda{\rm CDM})$, can be derived from constraints using LSS.
    Since each of these distributions are difficult to quantify in a simple way, any comparison between them is also complicated.
    A number of different measures are used to give a simple, generally ``single-numbered'', quantification of any differences~\cite{Inman, Bhattacharyya}, where~\cite{Marshall:2004zd, Verde:2013wza, Seehars:2014ora} are particularly used in cosmology and astronomy.
    The way each of these measures are interpreted can lead to confusing statements about any discordance and so a thorough discussion of a few of the major methods is laid out here.
    Detailed descriptions of each method, using some simple distributions, can be found in Appendix~\ref{Appendix} in order to help guide the reader.
    \\\\
    Consider the posterior distributions $P_1\equiv P(\btheta|D_1,{\rm M})$ and $P_2\equiv P(\btheta|D_2,{\rm M})$ for datasets $D_1$ and $D_2$, respectively, parameters, $\btheta$, of a model ${\rm M}$.
    \\\\
    \indent\emph{1.~Bhattacharyya~distance\label{p:b}} The Bhattacharyya distance~\cite{Bhattacharyya} compares the probability distributions from each model at a given parameter value
    \begin{equation}
        B = \int d\btheta\sqrt{P_1P_2}.\label{E:B}
    \end{equation}
    $B = 1$ indicates two identical distributions whilst $B\gtrsim0$ for disparate distributions with values in between indicating the level of tension.
    If one of the distributions is particularly broad compared to the other then this will give a low Bhattacharyya distance value meaning the distributions are distinctly different.
    This is true even if the peaks of the distributions are identical.
    The Bhattacharyya distance is not used in a cosmological context since the variance of the posterior distribution given LSS data is often much wider than when using measurements of the CMB.
    It is, however, easy to understand and aids in comprehension of comparisons between posterior distributions.
    \\\\
    \indent\emph{2.~Overlap~coefficient\label{p:o}} The overlap coefficient~\cite{Inman} works in a similar way to the Bhattacharyya distance.
    In this case the quantity obtained is given by
    \begin{equation}
        O = \int d\btheta{\rm Min}[P_1, P_2].\label{E:O}
    \end{equation}
    As with $B$, two identical distributions have $O=1$ and non-overlapping distributions have $O=0$.
    The scale of difference between $0<O<1$ is not the same as the Bhattacharyya distance, with the overlap coefficient taking lower values for the same pair of differing distributions.
    Again broader distributions are indicated as being in \emph{tension}, even with identical distribution peaks.
    This is also not often used for cosmological comparison.
    \\\\
    \indent\emph{3.~Difference~vector\label{p:c}} This measure, introduced in \cite{Battye:2014qga} and inspired by the two sample T-test~\cite{Rice:2006}, involves calculating the difference between the parameter ranges from the first and second probability distributions and creating a new probability distribution from the difference vector%
        \begin{equation}
            P(\dtheta|D_1,D_2,\mathcal{M}) =\int d\btheta' P_1(\btheta')P_2(\btheta'-\dtheta).\label{E:C}
        \end{equation}
        Here $\dtheta=\btheta_1-\btheta_2$, where $\btheta_1$ and $\btheta_2$ are the allowed values of the parameters from the distributions from data set $D_1$ and data set $D_2$, thus $P_2(\btheta_1-\dtheta)\equiv P_2(\btheta_2)$.
        This convolution effectively shifts the mean of the new distribution to the difference in the means of the original two distributions, $\mu_{\dtheta}=\mu_{\btheta_1}-\mu_{\btheta_2}$, with parameters spanning a range from $\mu_{\dtheta}-{\rm Min}[\btheta_1,\btheta_2]$ to $\mu_{\dtheta}+{\rm Max}[\btheta_1,\btheta_2]$. 
        For convenience $P(\dtheta|D_1,D_2,\mathcal{M})$ is denoted $P(\dtheta)$.
        A quantification of the disagreement between the distributions is obtained by integrating this new distribution within the isocontour formed by the value of the probability distribution function at $\dtheta={\bf 0}$,
        \begin{equation}
            C = \int_{A} d\dtheta P(\dtheta),\label{E:C}
        \end{equation}
        where
        \begin{equation}
            A = \left\{ \dtheta \left| P(\dtheta)>P({\bf 0})\right\}\right.
        \end{equation}
    In \cite{Battye:2014qga} samples were taken from MCMC chains and analysed, giving means for each parameter and a covariance matrix for each distribution.
    The covariance matrices were then combined using the law of total covariance~\cite{rudary2009predictive}. 
    This combined covariance was used to form a multivariate Gaussian distribution centred at the difference in the means of the parameters obtained from the {\ttfamily COSMOMC} analysis of the MCMC chains.
    In this paper the difference between the samples in the chains are used directly to form the probability distribution.
    This means that any non-Gaussianity of the distributions is taken into account.
    \\\\ 
    As a single unit measure this does a good job of indicating disagreements between distributions. 
    It can be interpreted easily since $C$ is a measure of the fraction of samples within a bounded area.
    This area is arbitrary and choosing $\dtheta={\bf 0}$ is not essential.
    Of course, the measure cannot fully describe the complexity of both of the entire probability distribution functions $P_1$ and $P_2$. 
    Using more parameters can help give greater understanding.
    \\\\
   \indent\emph{4.~Integration~between~intervals\label{p:ibi}} Using two numbers to quantify the similarities and differences between probability distributions \emph{can} provide more information.
    By integrating each of the probability distributions within a given interval of the other distribution, the total level of agreement can be quantified.
    The two useful numbers here are
    \begin{align}
        I_1 = &\phantom{a}\int_{A_2}d\btheta P_1\\
        I_2 = &\phantom{a}\int_{A_1}d\btheta P_2,
    \end{align}
    where 
    \begin{equation}
        A_i \equiv \left\{\btheta~\bigg|\int d\btheta P_i = 0.997\right\}.
    \end{equation}
    $I_1$ is obtained by integrating the probability distribution $P_1$ within the isocontour of the probability distribution $P_2$ which would contain 99.7\% of the samples drawn from it.
    $I_2$ is obtained in exactly the same way, exchanging the probability distribution $P_1$ for $P_2$.
    This measure is particularly useful since $I_1$ and $I_2$ can be directly related to samples obtained via MCMC analysis.
    The limit chosen for the integration interval is arbitrary.
    If the interval is chosen to measure the amount of $P_1$ within the isocontour which contains 68.4\% of $P_2$ then, if $I_1 = 0$, the \emph{tension} could be interpreted as being greater than $1\sigma$.
    We have chosen to consider isocontours containing 99.7\% of the samples from each distribution.
    If $I_1=0$ when integrated within the bounds containing 99.7\% of the samples drawn from $P_2$ then $P_1$ would be considered to be in $>3\sigma$ tension with $P_2$.
    Although computationally intensive, this method can be used to quantify an exact tension by increasing the integration limits of one of the distributions until the integral of the other distribution was no longer zero.
    This procedure is not performed in section~\ref{S:Discordance} due to computational resources.
    \\\\
    \indent\emph{5.~Surprise\label{p:s}} Another method which compares one distribution to another giving two measures is that used in \cite{Seehars:2014ora}.
    Here the relative entropy (KL-divergence) is found when $P_2$ is used as an update to $P_1$ and is given by
    \begin{equation}
        D(P_1||P_2) = \int d\btheta P_2\log\frac{P_2}{P_1}.\label{E:RelativeEntropy}
    \end{equation}
    An expected relative entropy can be found using
    \begin{equation}
        \langle D\rangle = \int dP_2\int d\btheta P_1P_2D(P_1||P_2).\label{E:ExpectedEntropy}
    \end{equation}
    By comparing the difference of the relative entropy to the expected relative entropy a quantity (which is named \emph{surprise} in \cite{Seehars:2014ora}) can be calculated
    \begin{equation}
        S = D(P_1||P_2) -\langle D\rangle.\label{E:Surprise}
    \end{equation}
    Using a combination of $D(P_1||P_2)$ and $S$ a quantification of information gain due to different distributions can be found.
    $S$ should be close to zero for datasets which are similar and can be positive or negative.
    A positive ``surprise'' indicates that the distribution used to update the original is more different than expected.
    A negative ``surprise'' is obtained when the updating distribution is in more agreement than expected with the original distribution.
    This technique is particularly useful when comparing the amount of ``surprise'' for a given expected relative entropy.
    The results of which can be quoted as a $p$-value and interpreted as how likely one distribution is to be in agreement with the other.
    \\\\
    \indent\emph{6.~Quantification~of~Bayesian~evidence\label{p:m}} Other measures that have previously been discussed generally involve comparisons of Bayesian evidences.
    The most simple and commonly used was introduced in \cite{Marshall:2004zd}.
    This is given by
    \begin{equation}
        R = \frac{p(D_1, D_2)}{p(D_1)p(D_2)},\label{E:R}
    \end{equation}
    where $p(D_i)$ is the evidence given data $D_i$,
    \begin{equation}
        p(D_i) = \int d\btheta P_ip(\btheta),
    \end{equation}
    where $p(\btheta)$ is the prior on the parameter $\btheta$ and the index $i=1,2$ denotes which dataset is used.
    The numerator of Eqn.~\ref{E:R} is given by
    \begin{equation}
        p(D_1, D_2) = \int d\btheta P_{1}P_2p(\btheta).
    \end{equation}
    This is related quite closely to the Bhattacharyya distance.
    $R$ is the ratio of the evidence given both datasets, to the evidence of each dataset.
    The prior assumptions of the parameter must be specified and taken into account.
    Using $\log R$, the results can be interpreted on the Jefferys scale with $\log R>0$ indicating agreement and $\log R<0$ indicating disagreement to some degree.
    This, as for the Bhattacharyya distance and the overlap coefficient methods, reveals a degeneracy between shifts in the peaks of distributions and broadening of the variances of distributions.
    The numbers from $\log R$ are dependent on the choice of priors.
    As long as the prior is stated along with analysis then the results can be recreated and interpreted by the individual. 
    \\\\
   \indent\emph{7.~Shifted~probability~distribution\label{p:v}} Another measure, used in~\cite{Verde:2013wza}, shifts one distribution (in a similar way to the difference vector method) so that the maxima of the two distributions coincide is then found.
    The ratio of the joint evidences is
    \begin{equation}
        T = \frac{p(D_1,D_2)_{\rm shifted}}{p(D_1,D_2)}.
    \end{equation}
    Identical distributions have $\log T=0$ and $\log T>0$ indicates deviations from similarity.
    The values of $\log T$ do not directly map to a statistical significance or a $p$-value.
    Also, $\log T$ can be expected to be twice as large when the dimensionality of the problem increases by two.
    This can either be corrected or taken into consideration when interpreting the result.
    \\\\    
    Each of the measures described in this section indicate, to some degree, whether or not two distributions agree or disagree with each other.
    They do not each give the same emphasis as to \emph{where} tension arises.
    \begin{itemize}
        \item{The Bhattacharyya distance, overlap coefficient and quantification of Bayesian evidence give disagreements arising from broadening of one distribution in comparison to another.
        The difference vector, shifted probability distribution, integration between intervals and ``surprise'' methods take this broadening into account.}
        \item{The Bhattacharyya distance and overlap coefficient have results which are difficult to interpret and do not map to any useful scales.}
        \item{The quantification of Bayesian evidence and shifted probability distribution methods are prior dependent and, out of the two, only $\log R$ can be interpreted on the Jeffreys scale.}
        \item{The ``surprise'' gives a variety of quantifications which can be mapped to two $p$-values, thus quantifying the amount of disagreement when either distribution is used to update the other.}
        \item{The difference vector relates the fraction of samples within an arbitrary boundary formed by the samples away from the difference in the means.
        It does not capture all the information, but can be quoted as a single number by mapping $C$ onto the interval of the 1D Gaussian.
        Due to its construction, the value of $C$ matches the expected results when comparing 2D likelihood contours, but extends to higher dimensions.}
        \item{Integration between intervals is more powerful than using $C$ for observing differences and it is easy to understand each integral individually.
        However, the combination needs to be taken into account to truly describe how much tension is present between distributions.
    This can lead to some confusion when considering a broad distribution compared with a tight one.}
    \end{itemize}
    In Sec.~\ref{S:Data}, the difference vector measure (\emph{3}) will be used for comparison of the constraints on {\greektext L}CDM parameters derived from the CMB and individual LSS probes.
    This represents an update of \cite{Battye:2014qga} on the basis of more recent data.
    In Sec.~\ref{S:Discordance}, each of the other statistics will be calculated in order to quantify the level of tension between the parameter distributions from the CMB to LSS with a discussion of what each implies.
    The probability distributions are complex and multi-dimensional, and so care needs to be taken when histogramming the samples from MCMC chains.
    These distributions can often be sparsely sampled in the important overlapping regions.
    For measures~\emph{1},~\emph{2} and~\emph{4}-\emph{7} the histogram is made for a number of different bins and both with and without Gaussian smoothing.
    The results quoted in Sec.~\ref{S:Discordance} are the consensus values from this range of tests (which are all quite similar in any case).
    For measure~\emph{3} the number of samples from the chains is much greater since there are $N_{\rm CMB}\times N_{\rm LSS}$ differences, where $N_{\rm CMB}$ is the number of samples from the CMB chains and $N_{\rm LSS}$ is the number of samples from the LSS chains.
    This is then histogrammed with a range of bins and with and without Gaussian smoothing to check that the results are robust.
	
	\section{Data\label{S:Data}}
	In the time since \cite{Battye:2014qga} there has been a number of data releases from CMB and LSS observations, as well as improved data analysis taking into account astrophysical uncertainty.
	Here, we consider updated versions of each of the data sets used in \cite{Battye:2014qga} to reassess the quantification of the tension in parameter constraints when obtained from the CMB and by LSS probes.

	\subsection{CMB}

\begin{figure}
		\centering
		\includegraphics{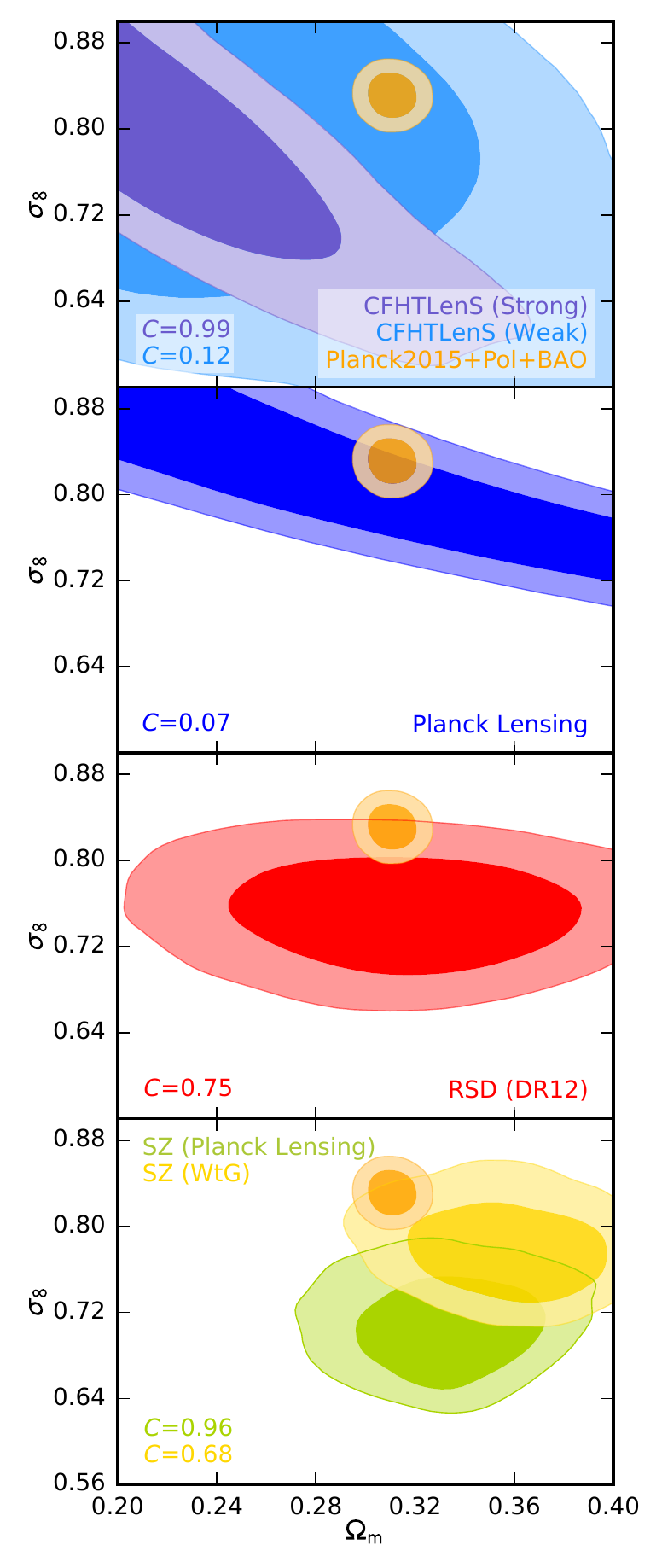}
		\caption{$1$ and $2\sigma$ constraint contours in the $\Omega_{\rm m}-\sigma_8$ plane within the {\greektext L}CDM model for a range of data.
		         In each subplot the orange contours show the constraints from \emph{Planck}2015+Pol+BAO.
		         The top subplot shows the constraints from weak lensing with the CFHTLenS \emph{(Strong)} and CFHTLenS \emph{(Weak)} results plotted in purple and and light blue respectively.
		         The second and third subplots show the constraint from CMB lensing in dark blue and from BOSS DR12 RSD in red.
		         The bottom subplot contains the constraint from SZ galaxy cluster counts with mass biases from CMB lensing in lime green and WtG in yellow.
		         The amount of disagreement between the combined five parameter constraints obtained by the CMB and probes of LSS are shown in the bottom-right of each subplot.}
		\label{fig:oms8}
	\end{figure}

	The temperature anisotropies and polarisation of photons from the CMB have been measured to an extremely high resolution over the largest possible scales by \emph{Planck}~\cite{Adam:2015rua}.
	For brevity, we consider only one combination of CMB measurements.\\\\
	\indent\emph{Planck2015+Pol+BAO}: 
	    We use the updated results from the \emph{Planck}2015 analysis. 
	    We include the temperature (T), E-mode and T-E cross-spectra from \emph{Planck} HFI for $29<\ell<2509$ and T, E- and B-mode spectra from \emph{Planck} LFI for $2<\ell<29$~\cite{Aghanim:2015xee}.
	    We combine this with the measure of the BAO peak from the 6dF Galaxy Survey~\cite{Beutler:2011hx} and the SDSS DR7 Main Galaxy Sample~\cite{Padmanabhan:2012hf} as in \cite{Battye:2014qga} and update the SDSS-III BOSS result to the final DR12 CMASS and LOWZ~\cite{Gil-Marin:2015nqa} result.
	    The \emph{Planck}2015+Pol+BAO $2\sigma$ constraint contours in the $\Omega_{\rm m}-\sigma_8$ plane can be seen in orange in Figs.~\ref{fig:oms8} and~\ref{fig:oms8_combined} as well as in Fig.~\ref{fig:oms8_neutrinos} where {\greektext L}CDM has been extended to include active or sterile neutrinos.
	    It should be noted that when quoting the discrepancy between results from the CMB and from LSS, it is the amount of disagreement in the five applicable {\greektext L}CDM parameters, not the tension in the two-dimensional $\Omega_{\rm m}-\sigma_8$ plane.
	
	\subsection{LSS}

	Large scale structure can be measured in a number of different ways.
	We consider four independent measurements of LSS which can be consistently combined to form a total constraint which we call \emph{All LSS}.
	Since LSS cannot constrain the optical depth to reionisation it is fixed to the central value from \emph{Planck}2015+Pol+BAO of $\tau=0.078$.
	For consistency with the thorough analysis of weak lensing using CFHTLenS~\cite{Joudaki:2016mvz} we adopt the same wide priors on the other {\greektext L}CDM parameters: $\Omega_{\rm b}h^2=[0.013,0.033]$; $\Omega_{\rm c}h^2=[0.01,0.99]$; $\theta_{\rm MC}=[0.5,10]$; $n_{\rm s} = [0.7,1.3]$; and $\log A_{\rm s}=[2.3,5]$.
	This is different to the prescription used previously in \cite{Battye:2014qga} which fixes tight, Gaussian priors to $\Theta_{\rm MC}$ and $n_{\rm s}$ which skews the other (somewhat-correlated) {\greektext L}CDM parameters to less favourable regions of parameter space. This can be considered as one of the reasons for the large apparent discordance between the CMB and LSS constraints.
	
	\subsubsection*{Galaxy lensing}

	Surveys of the gravitational lensing of photons by large scale structures are able to probe the underlying matter power spectrum of density perturbations and as such can give constraints on both $\Omega_{\rm m}$ and $\sigma_8$ directly~\cite{Kaiser:2000if}.
	The matter power spectrum arises from the correlation of the cosmological shear, but there are also other contributions from intrinsic alignments, i.e. shape-shape correlations due to galaxies forming near each other within the same gravitational potential or shape-shear correlations due to galaxies affecting the shear along the line of sight.
	These intrinsic alignment effects are small, but do contribute to the measurements from weak lensing surveys and should be taken into account.
	Modelling the gravitational lensing signature is also difficult since it involves knowing, to a high precision, galaxy dynamics.
	We present, here, three different constraints for weak lensing.\\\\
	\indent\emph{CFHTLenS~(Strong)}:
	    This relates directly to the \emph{Min} case in \cite{Joudaki:2016mvz} Fig.~12 which has the strongest assumptions made about astrophysical uncertainties.
	    There are seven angular bins and seven tomographic redshift bins which each have their own uncertainties related to them.
	    These redshift uncertainties are Gaussians about $\Delta z_1=-0.045\pm0.014$, $\Delta z_2 = -0.013\pm0.010$, $\Delta z_3=0.008,\pm0.008$, $\Delta z_4=0.042\pm0.017$ and $\Delta z_5=0.042\pm0.034$ leaving the last two bins with flat priors of $\Delta z_{6,7}=[-0.1,0.1]$, keeping all angular scales.
	    There are also tight priors on the amplitude of intrinsic alignments and the intrinsic alignment luminosity and redshift dependence are zero.
	    The $2\sigma$ constraint contours for CFHTLenS \emph{(Strong)} can be seen in the top subplot of Fig.~\ref{fig:oms8} in purple.
	    It can be seen that, in the $\Omega_{\rm m}-\sigma_8$ plane, the distribution lies far from the \emph{Planck}2015+Pol+BAO contours.
	    The value of $C = 0.99~(2.65\sigma)$.\\\\
	\indent\emph{CFHTLenS~(Weak)}:
	    As for the CFHTLenS \emph{(Strong)} case, this also comes from \cite{Joudaki:2016mvz} where it is denoted \emph{Max}.
	    The astrophysical assumptions are greatly reduced with wide flat priors on intrinsic alignment measurements and $\Delta z=[-0.1,0.1]$ for each of the seven tomographic bins, while non-linear scales are cut in the matter power spectrum.
	    The cut to the non-linear scales is the main cause for measurements from CFHTLenS \emph{(Weak)} being much less constraining than CFHTLenS \emph{(Strong)}.
	    This can also be found in the top subplot of Fig.~\ref{fig:oms8} in light blue.
	    Since the constraints are quite weak, there is clearly no discrepancy with \emph{Planck}2015+Pol+BAO in the $\Omega_{\rm m}-\sigma_8$ plane, although the central value is different.
	    The value of $C = 0.12~(0.15\sigma)$.\\\\
	\indent\emph{DES Science Verification}:
	    The results from the Dark Energy Survey (DES) follow the prescription in \cite{Abbott:2015swa} where the range of angular scales included is less than in either of the CFHTLenS analyses for each of its three redshift bins.
	    Here uncertainties in the redshift bins are not taken into account and intrinsic alignments are set to zero.
	    As such the constraints are not as tight as the CFHTLenS \emph{(Strong)} but provide a stronger constraint than CFHTLenS \emph{(Weak)}.
        Although not shown in Fig.~\ref{fig:oms8} for presentational reasons, the discrepancy between this data and \emph{Planck}2015+Pol+BAO is $C = 0.63~(0.90\sigma)$.
        \\\\
        Since we performed this analysis the Kilo-Degree Survey (KIDS)~\cite{Kuijken:2015vca} has produced results which are similar in many ways to those produced by CFHTLenS. Given this, we have not quoted a value for this data presuming it to be close to that for CFHTLenS.
	
	\subsubsection*{CMB lensing}

	Measuring the gravitational lensing of CMB photons can also provide information about cosmological shear correlations related to the matter power spectrum, hence revealing information about $\Omega_{\rm m}$ and $\sigma_8$~\cite{Lewis:2006fu}.\\\\
	\indent\emph{Planck~lensing}:
	    As well as measuring the primary anisotropies and polarisation, \emph{Planck} also detected the effects of the gravitational lensing of CMB photons.
	    Here, we use the measurements of the lensing power spectrum between $40<\ell<400$, as in \cite{Ade:2015zua}.
	    As expected, there is no discrepancy between \emph{Planck} lensing and the measurements of the CMB temperature and polarisation from \emph{Planck}2015+Pol+BAO. This can be seen in the $\Omega_{\rm m}-\sigma_8$ plane in the second subplot of Fig.~\ref{fig:oms8}, in particular we find that $C = 0.07~(0.08\sigma)$.

	\subsubsection*{Redshift-space distortions}

	Non-linear effects from the peculiar velocities of galaxies within galaxy clusters can be measured by surveys in redshift-space.
	In particular, finger-like structures can form in redshift-space due to the velocities of galaxies falling towards the centre of galaxy clusters.
	Measuring the deviation of observations from a fiducial cosmology allows the RSD to be quantified into the Alcock-Paczynski factor $F_{\rm AP}$, which is related directly to the Hubble parameter $H(z)$, and the angular diameter distance $D_{\rm A}(z)$.
	The joint growth of structure and amplitude of density perturbations of dark matter $f\sigma_8$, can also be constrained using the relative amplitudes of the RSD monopole and quadrupole~\cite{Beutler:2013yhm}.\\\\
	\indent\emph{SDSS-III BOSS DR12 RSD}: 
	    Measurements of the clustering of galaxies along the line of sight at effective redshifts of $z_{\rm LOWZ}=0.32$ and $z_{\rm CMASS}=0.57$ can constrain $f\sigma_8$ and the combination of the Hubble parameter, the comoving sound horizon at the baryon drag epoch $H(z)r_{\rm s}(z_{\rm d})$ and ratio of the angular diameter distance to the sound horizon $D_{\rm A}(z)/r_{\rm s}(z_{\rm d})$~\cite{Gil-Marin:2015sqa}.
	    Here, we use the covariance matrix for these parameters from the Quick-Particle-Mesh (QPM) mocks.
	    The constraints coming from RSD in the $\Omega_{\rm m}-\sigma_8$ plane can be seen in the third subplot of Fig.~\ref{fig:oms8} in red which suggests it is consistent with \emph{Planck}2015+Pol+BAO, but with the central value lying lower in the $\sigma_8$ direction.
	    In this case $C =0.75~(1.16\sigma)$. 
	
	\subsubsection*{Sunyaev-Zel'dovich galaxy cluster counts}

	Inverse Compton scattering of CMB photons by high energy electrons in intracluster media can be used to measure the number of galaxy clusters as a function of redshift, from which the growth of structure and various geometrical factors can be constrained~\cite{Ade:2015fva}.
	The relationship between the observable, $Y$, and the mass of the cluster, $M$, must be determined empirically using either observations or simulations.
	A simple assumption for the thermal state of a cluster is to assume hydrostatic equilibrium~\cite{Ade:2015fva}, and any deviation from the $Y-M$ relation derived from this assumption is quantified using a hydrostatic mass bias $1-b$.
    This factor can be constrained using follow-up observations of X-ray detected samples using weak lensing or directly from the lensing effect of clusters on the CMB measured from the \emph{Planck} data.\\\\
	\indent\emph{Planck lensing}:
	    The lensing effect of clusters on the CMB can be used to infer $1/(1-b)=0.99\pm0.19$~\cite{Ade:2015zua}.
	    Constraints using this mass bias are presented in lime green in the bottom subplot of Fig.~\ref{fig:oms8} for the $\Omega_{\rm m}-\sigma_8$ plane where it can be seen that the disagreement with the CMB is significant.
	    The value of $C = 0.96~(2.01\sigma)$.\\\\
	\indent\emph{Weighing the Giants (WtG)}: 
	    There are 51 galaxy clusters in the sample studied by the WtG project, 22 of which overlap with the \emph{Planck} galaxy clusters, for which lensing data exists~\cite{Mantz:2014paa}.
	    The mass bias determined by WtG is lower than for \emph{Planck} at $1-b=0.688\pm0.092$ and as such galaxy cluster dynamics suggest that these objects deviate significantly from hydrostatic equilibrium. 
	    The $1$ and $2\sigma$ constraint contours in the $\Omega_{\rm m}-\sigma_8$ plane can be found in the bottom subplot of Fig.~\ref{fig:oms8} in bright yellow, showing reasonable overlap with the \emph{Planck}2015+Pol+BAO constraints, such that $C = 0.68~(0.99\sigma)$.

	\section{Concordance or discordance?\label{S:Discordance}}
	
\begin{figure}
		\centering
		\includegraphics{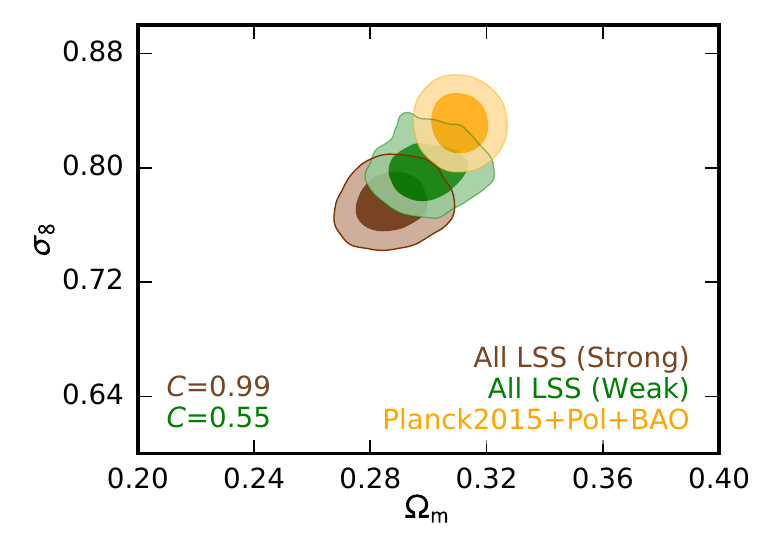}
		\caption{The $1$ and $2\sigma$ constraints on the $\Omega_{\rm m}-\sigma_8$ plane from \emph{Planck}2015+Pol+BAO in orange and from combining each of the LSS data sets, with those in the most tension with the CMB data set in brown and in the least tension in green.
		        The 5 parameter {\greektext L}CDM difference vector with \emph{Planck}2015+Pol+BAO is quoted for both sets of constraints in the bottom-left corner.}
		\label{fig:oms8_combined}
	\end{figure}

	Since each of the LSS probes are independent measurements, they can be combined to provide an \emph{All LSS} constraint.
	As pointed out in~\cite{Battye:2014qga}, if each of the mildly discrepant LSS constraints lie in the same region of parameter space, then their combination could become more significant than each separately.
	In order to investigate this we consider two combinations of data. \\\\
	\indent\emph{All LSS (Weak)}:
	    By combining CFHTLenS \emph{(Weak)} with \emph{Planck} lensing, RSD (DR12) and SZ galaxy cluster counts using the WtG mass bias we find the least discrepant joint analysis compared to \emph{Planck}2015+Pol+BAO. 
	    We can see in Fig.~\ref{fig:oms8_combined} the green contours in the $\Omega_{\rm m}-\sigma_8$ plane have reasonable overlap with \emph{Planck}2015+Pol+BAO and the value of $C= 0.55~(0.76\sigma)$.\\\\
	\indent\emph{All LSS (Strong)}: 
	    Combining the CFHTLenS \emph{(Strong)} constraints with \emph{Planck} lensing, RSD (DR12) and SZ galaxy cluster counts using the mass bias from \emph{Planck} lensing is shown in brown in Fig~\ref{fig:oms8_combined}.
	    This provides the most discrepant combination of data with $C = 0.99~(2.55\sigma)$.
	    Note that this is less discrepant than the CFHTLenS \emph{(Strong)} discrepancy by itself.
	    This suggests that there are internal tensions between the LSS data sets, as well as with CMB constraints.\\
    \begin{table}
        \centering
        \begin{tabular}{c|rcl|c}
        Measure & \multicolumn{3}{c|}{Result} & Interpretation \\\hline\hline&&&&\\
        \emph{1} & $B~$ & $=$ & $\phantom{-}1.81\times10^{-2}~$ & Unknown \\&&&&\\
        \emph{2} & $O~$ & $=$ & $\phantom{-}2.71\times10^{-3}$ & Unknown \\&&&&\\
        \emph{3} & $C~$ & $=$ & $\phantom{-}0.55~(0.76\sigma)$ & Low \\&&&&\\
        \multirow{2}{*}{\emph{4}} & $I_{\rm CMB}~$ & $=$ & $\phantom{-}3.81\times10^{-1}$ & \multirow{2}{*}{Low} \\
                 & $I_{\rm LSS}~$ & $=$ & $\phantom{-}2.30\times10^{-3}$\\&&&&\\
        \multirow{4}{*}{\emph{5}} & $~D({\rm CMB}||{\rm LSS})~$ & $=$ & $\phantom{-}7.20\times10^{-2}$ \\
        &$S_{{\rm CMB}\to{\rm LSS}}~$ & $=$ & $-4.25\times10^{-1}$ & Likely \\
        &$D({\rm LSS}||{\rm CMB})~$ & $=$ & $\phantom{-}8.52$ & similar\\
        &$S_{{\rm LSS}\to{\rm CMB}}~$ & $=$ & $\phantom{-}8.03$ \\&&&&\\
        \emph{6} & $\log R~$ & $=$ & $\phantom{-}3.29$ & Low \\&&&&\\
        \emph{7} & $\log T~$ & $=$ & $\phantom{-}2.59$ & Mild\\&&&&
        \end{tabular}
        \caption{Quantification of the similarity of the probability distributions of the {\greektext L}CDM parameters from \emph{Planck}2015+Pol+BAO and \emph{All LSS (Weak)} for each of the measures \emph{1}-\emph{7} from Sec.~\ref{S:Tension}.
                 The first column contains the measure used, the second column shows the result and the final column gives a description of degree of discordance.}
        \label{t:Comp_weak}
    \end{table}
        \begin{table}
        \centering
        \begin{tabular}{c|rcl|c}
        Measure & \multicolumn{3}{c|}{Result} & Interpretation \\\hline\hline&&&&\\
        \emph{1} & $B~$ & $=$ & $\phantom{-}8.90\times10^{-4}$ & Unknown \\&&&&\\
        \emph{2} & $O~$ & $=$ & $\phantom{-}9.70\times10^{-5}$ & Unknown \\&&&&\\
        \emph{3} & $C~$ & $=$ & $\phantom{-}0.99~(2.55\sigma)$ & Moderate\\&&&&\\
        \multirow{2}{*}{\emph{4}} & $I_{\rm CMB}~$ & $=$ & $\phantom{-}2.82\times10^{-2}~$ & \multirow{2}{*}{Moderate} \\
                 & $I_{\rm LSS}~$ & $=$ & $\phantom{-}5.44\times10^{-5}$\\&&&&\\
        \multirow{4}{*}{\emph{5}} & $~D({\rm CMB}||{\rm LSS})~$ & $=$ & $\phantom{-}2.85\times10^{-3}$ \\
        &$S_{{\rm CMB}\to{\rm LSS}}~$ & $=$ & $-5.85$ & Likely\\
        &$D({\rm LSS}||{\rm CMB})~$ & $=$ & $\phantom{-}7.84$ & different\\
        &$S_{{\rm LSS}\to{\rm CMB}}~$ & $=$ & $\phantom{-}1.99$ \\&&&&\\
        \emph{6} & $\log R~$ & $=$ & $-1.36$ & Significant \\&&&&\\
        \emph{7} & $\log T~$ & $=$ & $\phantom{-}7.56$ & Significant\\&&&&
        \end{tabular}
        \caption{Identical table to Table.~\ref{t:Comp_weak} using \emph{All LSS (Strong)} to constrain the LSS parameter distributions.
                 The first column contains the measure used, the second column shows the result and the final column gives a description of the degree of discordance.}
        \label{t:Comp_strong}
    \end{table}\\
    In Tables.~\ref{t:Comp_weak} and~\ref{t:Comp_strong} we present the calculated values for each of the statistics. 
    The overall picture is that \emph{All LSS (Strong)} is more discrepant than the parameter distributions inferred from \emph{Planck}2015+Pol+BAO while using \emph{All LSS (Weak)} appears to be more compatible.
    However, the details indicate a more complicated story dependent on which measure is used.
    \\
    \\
   The results of measures \emph{1} and \emph{2} in Tables.~\ref{t:Comp_weak} and~\ref{t:Comp_strong} are small compared to $B=1$ or $O=1$ suggesting a large degrees of discordance between constraints obtained from LSS and CMB datasets.
   To illuminate how poor these measures are at quantifying tension, a toy model can be considered to see what the results are equivalent to in terms of shifts of two distributions.
   If $P_1=\mathcal{N}({\boldsymbol\mu}_1,{\bf \Sigma})$ and $P_2=\mathcal{N}({\boldsymbol\mu}_2,{\bf\Sigma})$ with ${\boldsymbol\mu}_1=(0,0,0,0,0)$, ${\boldsymbol\mu}_2=(0,0,0,0,\theta)$ and ${\bf\Sigma}={\rm diag}(1,1,1,1,1)$ then $B=1.81\times10^{-2}$ needs $\theta=4.23$ whilst $B=8.90\times10^{-4}$ needs $\theta=5.59$.
   In a similar way $O=2.71\times10^{-3}$ requires $\theta=3.48$ and $O=9.70\times10^{-5}$ needs $\theta=4.42$.
   From these shifts in the five dimensional distributions it appears that \emph{All LSS (Weak)} and \emph{All LSS (Strong)} are both quite distinct from \emph{Planck}2015+Pol+BAO.
    There is a strong dimensional dependence using these two measures so extremely small values can, and do, appear as large discrepancies.
    On the basis of this, these measures indicate significant discordance between \emph{All LSS (Weak)} and \emph{Planck}2015+Pol+BAO and severe discordance between \emph{All LSS (Strong)} and \emph{Planck}2015+Pol+BAO.
    However, since the shift in the means has an equivalent description in terms of broadening of the variance then it is difficult to make any useful statement.
    Instead consider another toy model where $P_1=\mathcal{N}({\boldsymbol\mu},{\bf \Sigma}_1)$ and $P_2=\mathcal{N}({\boldsymbol\mu},{\bf\Sigma}_2)$ with ${\boldsymbol\mu}=(0,0,0,0,0)$, ${\bf\Sigma}_1={\rm diag}(1,1,1,1,1)$ and ${\bf\Sigma}_2={\rm diag}(\sigma^2,\sigma^2,\sigma^2,\sigma^2,\sigma^2)$ then $B=1.81\times10^{-2}$ needs $\sigma\approx10$ whilst $B=8.90\times10^{-4}$ needs $\sigma\approx33$.
    Neither of these $P_2$ distributions would be considered in tension with $P_1$, although $P_2$ would not be informative.
    In general, the values of $B$ and $O$ are much less than one, which would suggest that there is reasonably significant discordance between \emph{All LSS (Weak)} or \emph{All LSS (Strong)} and \emph{Planck}2015+Pol+BAO.
    It is clear why neither the Bhattacharyya distance nor the overlap coefficient measures are used for data comparison in cosmology.
   \\
   \\
   Measure \emph{3} is easy to interpret in both Tables.~\ref{t:Comp_weak} and~\ref{t:Comp_strong}. 
   Since the value of $C$ is the fraction of samples within an interval then this maps easily to the number of samples in an interval of a 1D Gaussian distribution.
   This means that $C$ maps directly to a quantification in terms of a number of standard deviations.
   For \emph{All LSS (Weak)} compared to \emph{Planck}2015+Pol+BAO, $C=0.55$ is equivalent to $0.76\sigma$, which is interpreted as very little discordance.
   Comparing \emph{All LSS (Strong)} to \emph{Planck}2015+Pol+BAO provides $C=0.99$ which (including more significant figures in the calculation) maps to $2.55\sigma$.
   While this is much greater than for \emph{All LSS (Weak)}, the suggested interpretation of the tension is only a moderate one.
   These values reflect the position of the contours in Fig.~\ref{fig:oms8_combined}.
   \\
   \\
   When interpreting measure \emph{4} it is only necessary to consider ${\rm Max}[I_{\rm CMB},I_{\rm LSS}]$ to get an indication of the level of agreement.
   The ratio of the larger value to the smaller value then describes the relative widths of the distributions.
   In a similar way to measure \emph{3}, the values of $I_{\rm CMB}$ and $I_{\rm LSS}$ relate directly to numbers of samples, although the distributions are cut at the complementary distributions $3\sigma$ isocontours, meaning they are discontinuous.
   While this means they cannot truly be mapped to intervals of a 1D Gaussian, that is still a useful indicator of discordance.
   For \emph{All LSS (Weak)} 38.1\% of the samples drawn from the \emph{Planck}2015+Pol+BAO distribution are within the isocontour which would contain 99.7\% of the samples drawn from the \emph{All LSS (Weak)} distribution.
   This seems like a small fraction, but is actually the equivalent of a discrepancy of $0.88\sigma$ when compared to a 1D Gaussian and so should be interpreted as indicating a low level of discordance.
   Since $I_{\rm CMB}>I_{\rm LSS}$ then the constraints on the parameters using \emph{Planck}2015+Pol+BAO are tighter than those from \emph{All LSS (Weak)}.
   Similarly, $I_{\rm CMB}$ is larger in Table.~\ref{t:Comp_strong} showing that the constraints from \emph{Planck}2015+Pol+BAO are tighter than those from \emph{All LSS (Strong)}.
   $I_{\rm CMB}=2.82\times10^{-2}$ means that 2.82\% of the samples drawn from the \emph{Planck}2015+Pol+BAO distribution are within the isocontour which would contain 99.7\% of the samples drawn from the \emph{All LSS (Strong)} distribution.
   Again, this seems quite low but is the equivalent to $2.2\sigma$ and so is again only moderately discordant.
   \\
   \\
   Measure \emph{5}, is a bit more difficult to interpret in both Tables.~\ref{t:Comp_weak} and~\ref{t:Comp_strong}.
   In the case of updating both the \emph{All LSS (Weak)} and the \emph{All LSS (Strong)} constraints with the constraints from \emph{Planck}2015+Pol+BAO there is little relative entropy, but have large negative ``surprise''.
   Since the values of the ``surprise'' are negative this suggests that the distributions are more similar than expected.
   It should be noted that this does not mean that the distributions are that similar, just that there is less of an information gain than expected.
   Indeed, it is is very difficult to quantify quite how severe the discordance is using this measure; it should rather be used to describe whether one dataset is likely to update another dataset. 
   The important outcome of the measure \emph{5} results from Tables.~\ref{t:Comp_weak} and~\ref{t:Comp_strong} is the similarity between the results for $D($\emph{All LSS (Weak)}$||$\emph{Planck}2015+Pol+BAO$)$ and those for $S_{{\rm{\it All~LSS~(Weak)\to Planck}2015+Pol+BAO}}$.
   This indicates that the distributions are likely to be similar, whereas $D($\emph{All LSS (Strong)}$||$\emph{Planck}2015+Pol+BAO$)$ being larger than $S_{{\rm{\it All~LSS~(Strong)\to Planck}2015+Pol+BAO}}$ shows that it is more probable that the parameter distributions from \emph{Planck}2015+Pol+BAO can be updated with the constraints from \emph{All LSS (Strong)}.
   This means the distributions are likely to be more distinct.
   \\
   \\
   For measure \emph{6}, Table.~\ref{t:Comp_weak} has $\log R=3.29$ signifying that the joint distribution with \emph{Planck}2015+Pol+BAO and \emph{All LSS (Weak)} as data sets is more likely than each of the distributions separately. The similarity is quite significant when using flat priors from the minimum to maximum parameter values obtained in the samples.
    This happens only when the two distributions are at worst mildly discordant.
    When comparing this to the \emph{All LSS (Strong)} result of $\log R=-1.36$, in Table.~\ref{t:Comp_strong}, the negative value shows that the joint distribution is less likely than each of the distributions separately, which is true when the distributions are more distinct.
    It is best to interpret the values of $\log R$ on the Jeffreys scale often used in Bayesian analysis~\cite{Jeffreys:1961}, with a result of $\log R=3.29$ showing \emph{Planck}2015+Pol+BAO is ``decisively similar'' to \emph{All LSS (Weak)} and $\log R=-1.36$ suggesting \emph{Planck}2015+Pol+BAO is significantly different to \emph{All LSS (Strong)}.
    These statements are more extreme than the other measures as a result of placing relatively tight priors.
    Increasing the range of the prior distribution allows less extreme interpretation of the results but with the same quantitative outcome - the \emph{All LSS (Weak)} distribution is more similar to \emph{Planck}2015+Pol+BAO than the \emph{All LSS (Strong)} distribution is.
    \\
    \\
    Finally, measure \emph{7} indicates that the discordance between \emph{All LSS (Weak)} and \emph{Planck}2015+Pol+BAO is mild, but as with measure \emph{6} this statement is prior dependent.
    Again, the $\log T$ value when using \emph{All LSS (Strong)} gives a much more significant discordance.
    By changing the priors, the interpretation of this result can change from \emph{All LSS (Weak)} being in almost complete agreement with \emph{Planck}2015+Pol+BAO to there being significant or severe disagreement.
    The interpretation from \emph{All LSS (Strong)} then follows suit, being always more discordant than \emph{All LSS (Weak)}.
    \\
    \\
    To summarise the usefulness of each of these methods:
    \begin{itemize}
        \item{\emph{1} and \emph{2} cannot give a useful quantification of discordance, although the small values would \emph{suggest} more significant discordance than other methods.}
        \item{\emph{3} and \emph{4} can be related to drawn samples from distributions and so mapped to intervals on a 1D Gaussian and tend to give slightly more conservative interpretations of the discordance.}
        \item{\emph{5} is useful to find out whether a distribution is likely to usefully update the distribution from a pre-existing dataset, but cannot be easily interpreted as a quantification of the difference between the datasets.}
        \item{\emph{6} and \emph{7} are prior dependent and so care needs to taken when interpreting the actual values as a indication of the severity of discordance.}
    \end{itemize}
    
	\noindent All of these measures, for both \emph{All LSS (Weak)} and \emph{All LSS (Strong)}, are not representative of the value of the tension obtained in \cite{Battye:2014qga}. 
	This is true even though the $\Omega_{\rm m}-\sigma_8$ $2\sigma$ contour for the \emph{All LSS (Strong)} looks similar to the contour in the left hand subplot of Fig.~3 in \cite{Battye:2014qga}.
	This is due to updated data and the application of \emph{Planck}2013 priors on $\Theta_{\rm MC}$ and $n_{\rm s}$ in~\cite{Battye:2014qga}.
	These parameters were chosen since they are well measured by the CMB and in particular $\Theta_{\rm MC}$ is known to within 0.05\%. 
	\begin{figure}
	\centering
	\includegraphics[width=0.47\textwidth]{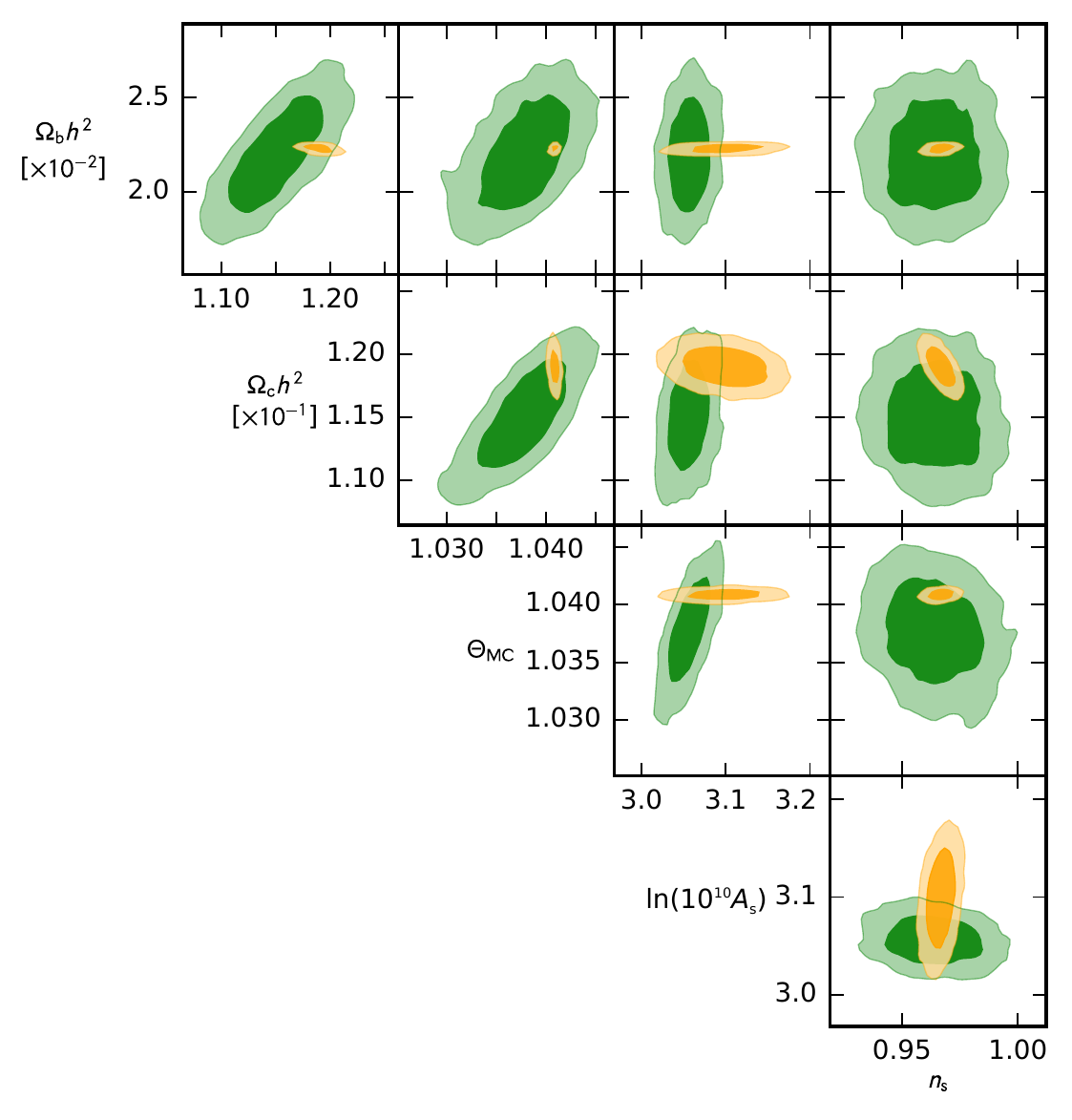}
	\vskip-1em
	\caption{Projected $1$ and $2$ $\sigma$ likelihood contours for each of the relevant {\greektext L}CDM parameters.
	The green contours show the \emph{All LSS (Weak)} constraints and the orange contours are the constraints from \emph{Planck}2015+Pol+BAO.}
	\label{fig:E2}
	\end{figure}
	\begin{figure}
	\centering
	\includegraphics[width=0.47\textwidth]{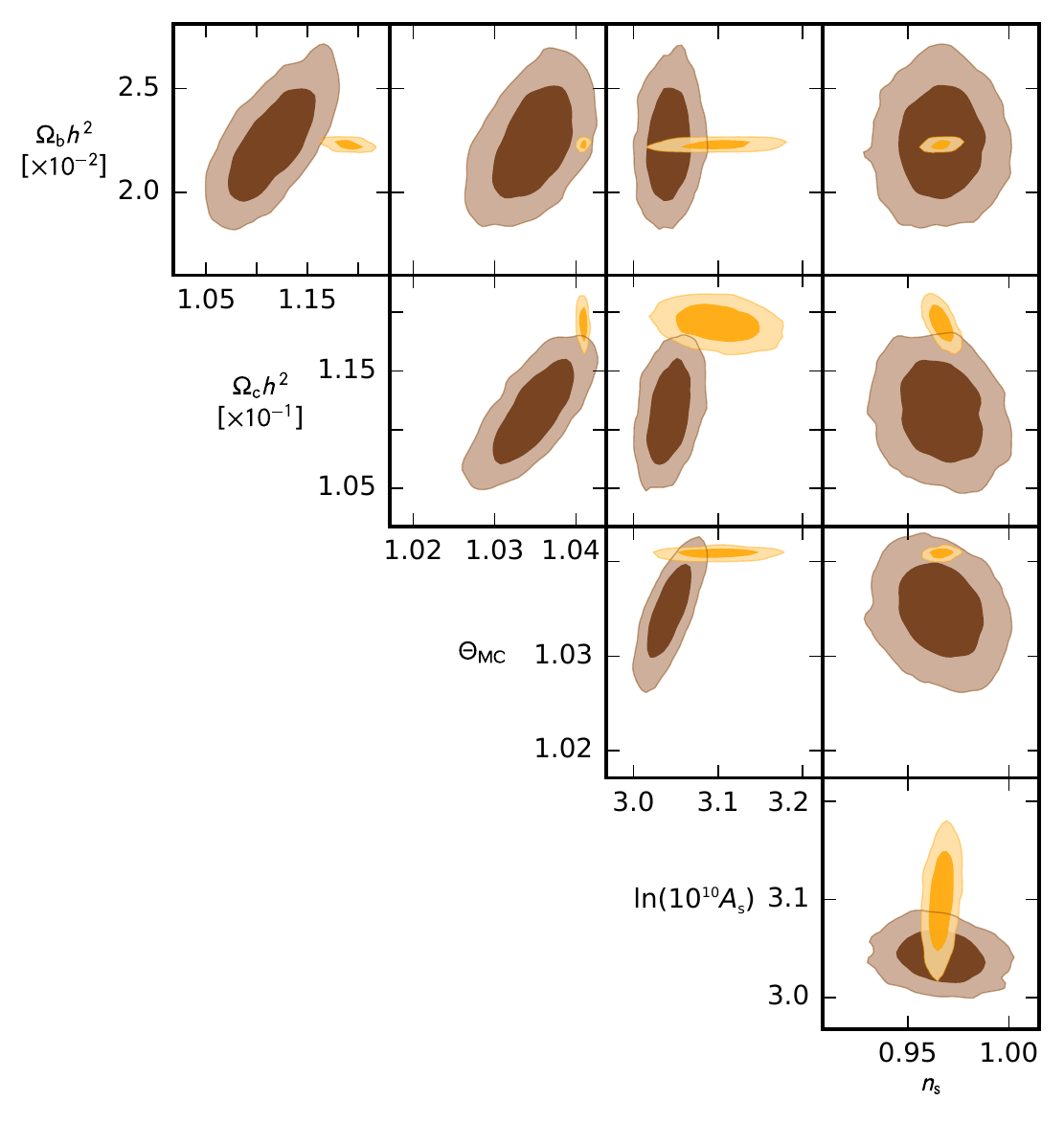}
	\vskip-1em
	\caption{Projected $1$ and $2$ $\sigma$ likelihood contours for each of the relevant {\greektext L}CDM parameters.
	The brown contours show the \emph{All LSS (Strong)} constraints and the orange contours are the constraints from \emph{Planck}2015+Pol+BAO.}
	\label{fig:E1}
	\end{figure}%
	Using importance sampling on the \emph{All LSS (Strong)} chains and placing priors of $\Theta_{\rm MC}=1.04086\pm0.00048$ and $n_{\rm s}=0.9652\pm0.0062$, the resulting tension with \emph{Planck}2015+Pol+BAO is $C=0.999(95)~(4.06\sigma)$ which is in closer agreement with the previously found result.
	There are relatively few samples in the prior regions of $\Theta_{\rm MC}$ and $n_{\rm s}$ when using \emph{All LSS (Strong)} as can be seen in certain regions of Fig.~\ref{fig:E1}.
	This means that the probability distribution from the samples is likely not to be representative of the true distribution.
	\\
	\\
	Since these values are restricted to a particular region of their parameter space, the other three {\greektext L}CDM parameters ($\Omega_{\rm b}h^2$, $\Omega_{\rm c}h^2$ and $\log A_{\rm s}$) become constrained to less favourable regions.
	Figs.~\ref{fig:E2} and ~\ref{fig:E1} show the projected likelihood contours comparing \emph{Planck}2015+Pol+BAO to \emph{All LSS (Weak)} and \emph{All LSS (Strong)} respectively.
	Although, not entirely accurate - the application of priors on $\Theta_{\rm MC}$ and $n_{\rm s}$ would restrict the green and the brown contours to the size of the \emph{Planck}2015+Pol+BAO contours in the $\Theta_{\rm MC}$ and $n_{\rm s}$ directions.
	For Fig.~\ref{fig:E2}, even though the priors limit the \emph{All LSS (Weak)} parameter distributions in all directions, they do not become significantly more discrepant with \emph{Planck}2015+Pol+BAO.
	On the other hand, for Fig.~\ref{fig:E1}, the priors restrict $\log A_{\rm s}$ and $\Omega_{\rm c}h^2$ to the upper range of their allowed values. 	
	This causes a knock on effect requiring both lower and higher $\Omega_{\rm b}h^2$ values from the correlation with $\Omega_{\rm c}h^2$ and the allowed region from the priors respectively.
	This ``new constraint'' lies further from \emph{Planck}2015+Pol+BAO and so the agreement with \emph{All LSS (Strong)} with priors decreases.
	It should be noted that it is na{\"i}ve to use the combinations of the 2D contours in Figs.~\ref{fig:E2} and~\ref{fig:E1} to make serious assumptions about shifts in the distributions with the application of priors.
	The true distributions are five dimensional and can only be projected down to the 2D contours via marginalising out other parameters, therefore losing a lot of information in the process. 
	Releasing these priors to cover a wider range allows more natural values in the remaining parameters to be chosen.
	These new parameter values lie closer to the values from \emph{Planck}2015+Pol+BAO, reducing the tension with \emph{All LSS}.
	\\
	\\
	It should be noted that if the belief in the $n_{\rm s}$ and $\Theta_{\rm MC}$ priors is strong, the result in greater tension may be favoured.
	We do not consider the application of these priors further since we give no preference to which data is used to constrain well understood cosmology.
	
	\section{Discussion\label{Disc}}
	\subsection{Neutrinos}
	
\begin{figure}
		\centering
		\includegraphics{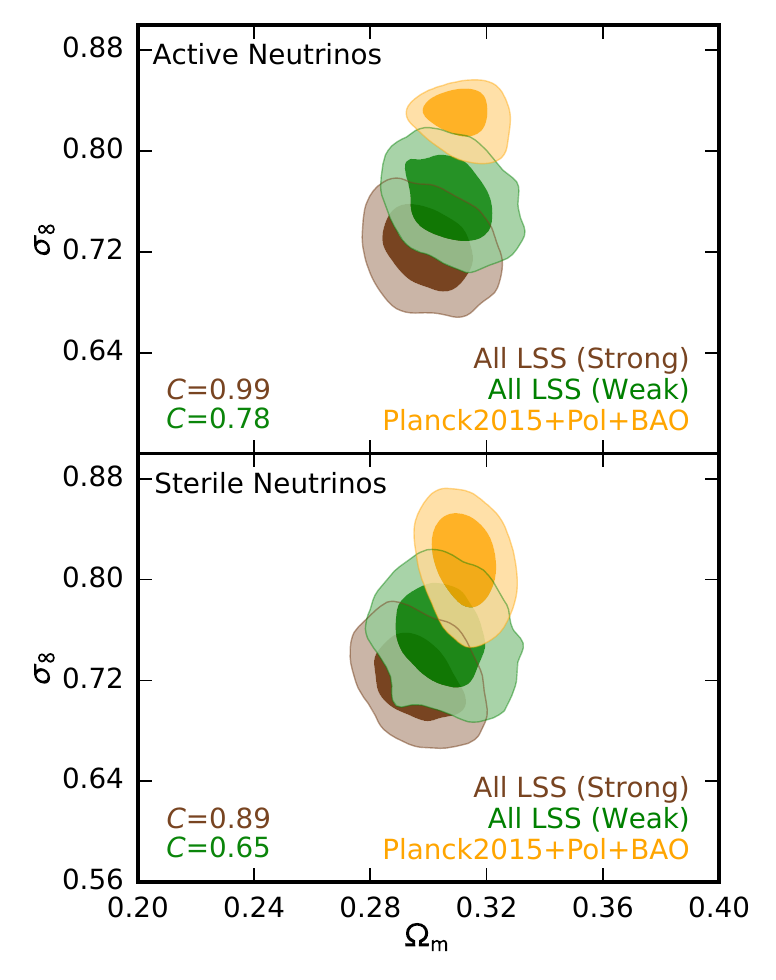}
		\caption{The $1$~and~$2\sigma$ $\Omega_{\rm m}-\sigma_8$ where {\greektext L}CDM has been extended by the inclusion of active neutrinos in the top subplot and sterile neutrinos in the bottom. Constraints from \emph{Planck}2015+Pol+BAO are in orange whilst constraints from the \emph{All LSS (Strong)} and \emph{All LSS (Weak)} combinations of LSS probes are in brown and green respectively. 
		The 5 parameter {\greektext L}CDM difference vector with \emph{Planck}2015+Pol+BAO is quoted for both sets of constraints in the bottom-left corner.}
		\label{fig:oms8_neutrinos}
	\end{figure}

	The inclusion of massive neutrinos into the model reduces the amount of small-scale power. This is because neutrino perturbations with Fourier modes longer than their comoving free-streaming length cannot cluster until these modes leave the comoving horizon. This occurs earlier in the Universe for more massive species~\cite{Bond:1980,Hu:1997,Elgaroy:2004,Lesgourgues:2006,Hannestad:2010,Hall:2012}.
	We consider the effects of both active and sterile neutrinos, i.e. three neutrinos with a combined mass of $\sum m_\nu$ equally divided between the three, and an additional neutrino with an effective mass $m_{\rm sterile}^{\rm eff}$ related to the true mass via the effective number of relativistic degrees of freedom $N_{\rm eff}$. 
	Using \emph{Planck}2015+Pol+BAO to constrain {\greektext L}CDM with either active or sterile neutrinos added indicates no preference for either a mass for active neutrinos, $\sum m_\nu<0.15{\rm eV}$ or any mass deviation of the number of relativistic degrees of freedom from sterile neutrinos, $m_{\rm sterile}^{\rm eff}<0.65{\rm eV}$ and $N_{\rm eff}<3.384$.
	\\\\
	Although the significance of the tension between constraints on {\greektext L}CDM parameters from the CMB and LSS \emph{without} neutrinos is reduced from the analysis in \cite{Battye:2014qga}, we can still investigate if the inclusion of neutrinos makes any additional difference. 
	Most importantly, we can see if there is any preference for massive neutrinos when combining CMB measurements with LSS observations. 
	The equivalent plot to Fig.~\ref{fig:oms8_combined} is shown in the upper subplot of Fig.~\ref{fig:oms8_neutrinos} when active neutrinos are included and in the lower subplot when sterile neutrinos are included.
	 \noindent Clearly there is very little benefit from adding active neutrinos, evident from the $\Omega_{\rm m}-\sigma_8$ contours.
	 When including $\sum m_\nu$, such that the probability distribution is six dimensional, $C=0.999(85)~(3.79\sigma)$ and $C=0.781~(1.23\sigma)$ for \emph{All LSS (Strong)} and \emph{All LSS (Weak)} respectively.
	 The distributions are marginally more discrepant with the addition of $\sum m_\nu$.
	 This is due to the distribution of neutrino mass not aligning particularly well in \emph{All LSS (Strong)} or \emph{(Weak)} compared to \emph{Planck}2015+Pol+BAO and not because an extra degree of freedom has been added.
	 Opposite to the result obtained in~\cite{Battye:2014qga}, the tension increases with the addition of active neutrinos.
     With the application of $\Theta_{\rm MC}$ and $n_{\rm s}$ priors, the $\sum m_{\nu}$ distributions aligned better along correlated parameter directions than when the priors are removed.
	 The value of the combined mass of neutrinos using \emph{All LSS (Strong)} or \emph{All LSS (Weak)} combinations with \emph{Planck}2015+Pol included becomes $\sum m_\nu=(0.176\pm0.056){\rm eV}$ or $\sum m_\nu=(0.146\pm0.057){\rm eV}$, where the significance of neutrino masses has slightly reduced from the $\sum m_\nu=(0.357\pm0.099){\rm eV}$ from \cite{Battye:2014qga}. 
	 The inclusion of this neutrino content does not help alleviate any tension between constraints from the CMB and LSS probes. 
	 This indicates that using active neutrinos as an extension to {\greektext L}CDM is not particularly useful, as suggested in \cite{Boris}.
	\\\\
	Sterile neutrinos fare a little better than their active counterparts in reducing the tension. 
	The visible overlap is slightly better than vanilla {\greektext L}CDM and much better than when active neutrinos are added.
	When we include these two parameters in the quantification analysis $C=0.891~(1.60\sigma)$ and $C=0.652~(0.94\sigma)$ when comparing \emph{Planck}2015+Pol+BAO to \emph{All LSS (Strong)} and \emph{All LSS (Weak)} respectively.
	This is in good agreement with what would be expected from the visual inspection of $\Omega_{\rm m}-\sigma_8$ contours.
	Due to the high dimensionality of this problem each bin in the histogram for $m_{\rm sterile}^{\rm eff}$ and $N_{\rm eff}$ is computed separately, written to disk and then analysed from the disk.
	This increases computation times significantly, especially when testing for a range of bin sizes and amounts of Gaussian smoothing.
	\\
	\\
	The values of $m_{\rm sterile}^{\rm eff}=(0.470\pm0.227){\rm eV}$ and $N_{\rm eff}=3.139\pm0.057$ or $m_{\rm sterile}^{\rm eff}=(0.234\pm0.115){\rm eV}$ and $N_{\rm eff}=3.162\pm0.059$ are obtained by combining \emph{All LSS (Strong)} or \emph{All LSS (Weak)} combinations with \emph{Planck}2015+Pol. These constraints are similar to the values expected from \emph{Planck}2015+Pol+BAO, although with peaks in their respective distributions.
	
	\subsection{\emph{Planck} 2016 results}
	
	In~\cite{Adam:2016hgk} the \emph{Planck}2015 temperature anisotropies are combined with the low-$\ell$ EE polarisation data ({\ttfamily lollipop}) from the \emph{Planck} high frequency instrument (HFI) and obtain a lower value of the optical depth to reionisation $\tau$.
	This shifts the \emph{Planck}2015 value of $\tau=0.078\pm0.019$ to $\tau=0.058\pm0.012$.
	It was suggested in ~\cite{Battye:2014qga} that constraining $\tau$ using LSS rather than the WMAP polarisation (which requries $\tau=0.091 \pm 0.013$) leads to a constraint of $\tau=0.049\pm0.021$ which is close to the new \emph{Planck}2016 value.\\\\
	The discrepancy between the values of $\tau$ inferred from WP and LSS suggested a possible resolution to the source of tension.
	With the new, lower constraint on $\tau$ from \emph{Planck}, the tension would be expected to be reduced.
	Since the \emph{Planck}+{\ttfamily lollipop} chains and likelihood code were not publicly available at present the \emph{Planck}2015+Pol+BAO chains were importance sampled using $\tau=0.058\pm0.012$.
	In this case, the quantification of tension when comparing to \emph{All LSS (Weak)} reduces from $C=0.550~(0.76\sigma)$ to $C=0.432~(0.57\sigma)$.
	This also reduces to a minor extent from $C=0.989~(2.55\sigma)$ to $C=0.985~(2.44\sigma)$ for the comparison to \emph{All LSS (Strong)}.
	At this stage we are unable to make any conclusive statement that the lowering of $\tau$ is in any more or less tension than \emph{Planck}2015+Pol+BAO.

	\section{Conclusions}	
	
    In this article, we have discussed a variety of measures with which to quantify the amount of discordance between any two probability distributions.
    By comparing methods, an understanding of how the different quantifications can be interpreted is found.
    Further, we have presented two new methods which are extremely robust and have an easy interpretation.
    The main point that we have made is that there are many issues arising from subjective interpretations of discordance.
    \\
    \\
    We have used the measure introduced in~\cite{Battye:2014qga} and described in detail here how to quantify the differences in the 5D {\greektext L}CDM parameter distributions when obtained by \emph{Planck}2015+Pol+BAO and a range of large scale structure probes.
    This update to~\cite{Battye:2014qga} is performed for different analyses of the same probe (SZ galaxy cluster counts from \emph{Planck} and Weighing the Giants for example) to show that the choice of analysis can significantly affect the constraints on parameters.
    By combining the LSS datasets in most tension with the CMB into \emph{All LSS (Strong)} and in least tension into \emph{All LSS (Weak)} two interpretations of the discordance are possible.
    In the \emph{All LSS (Strong)} case the discrepancy between the parameters is $C=0.989$ (around $2.55\sigma$ when mapping $C$ to an interval on a 1D Gaussian) which is greatly reduced from the value quoted in~\cite{Battye:2014qga}.
    Further analysis, imposing much tighter priors on $\Theta_{\rm MC}$ and $n_{\rm s}$ through importance sampling shows that the parameter distributions can be squashed into less likely regions, providing much greater discrepancy - in better agreement with the value in~\cite{Battye:2014qga}.
    This highlights how important it is to understand the applied priors and how they can affect the probability distribution.
    Further, if the belief in these priors is great enough then the tension remains problematic and more study into the alleviation of the discrepancy should again be considered.
    In the \emph{All LSS (Weak)} case, the difference from \emph{Planck}2015+Pol+BAO is almost non-existent at $C=0.550$ (around $0.76\sigma$).
    Each of the other measures discussed in this paper show the same general trend, that \emph{All LSS (Strong)} is more distant from \emph{Planck}2015+Pol+BAO than \emph{All LSS (Weak)}, but importantly, suggests the discordance exists to different degrees for each measure.
    \\
    \\
    We finished by discussing neutrinos, added as an extension to {\greektext L}CDM in~\cite{Battye:2013, Battye:2014qga}.
    We now conclude that active neutrinos provide no improvement over vanilla {\greektext L}CDM, worsening the discordance marginally with the \emph{All LSS (Strong)} combination and remaining similar for \emph{All LSS (Weak)}.
    Sterile neutrinos are somewhat better, reducing the parameter discrepancy marginally in the \emph{All LSS (Strong)} case when taking into account all seven of the relevant parameters.
    There is a slight worsening for \emph{All LSS (Weak)}, but only by a small amount.
    With the small discordance between the {\greektext L}CDM parameters without neutrinos, it is unlikely that their addition would be deemed necessary to solve the issue.
    \\
    \\
    We also discuss the implication of the \emph{Planck}+{\ttfamily lollipop} result.
    It was shown in~\cite{Battye:2014qga} that the optical depth to reionisation $\tau$, would be much lower if combining \emph{Planck} without polarisation results but instead in combination with LSS probes.
    This lower $\tau$ was then found when considering the low-$\ell$ EE polarisation from \emph{Planck}~\cite{Adam:2016hgk}.
    Attempting to mimic the \emph{Planck}+{\ttfamily lollipop} likelihood by imposing a range of priors, we find a very small reduction in the amount of tension between parameters obtained from CMB and LSS data.
    Since these chains do not contain the true probability distribution of the new \emph{Planck} results the reduction is not definite, but provide hints of an indication.

	\section*{Acknowledgments}
	TC is supported by an STFC studentship. AM is supported by a Royal Society University Research Fellowship. We are grateful for access to the University of Nottingham High Performance Computing Facility.
    
    \bibliography{cmblss2}

\def\eprinttmppp@#1arXiv:@{#1}
\providecommand{\arxivlink[1]}{\href{http://arxiv.org/abs/#1}{arXiv:#1}}
\def\eprinttmp@#1arXiv:#2 [#3]#4@{\ifthenelse{\equal{#3}{x}}{\ifthenelse{
\equal{#1}{}}{\arxivlink{\eprinttmppp@#2@}}{\arxivlink{#1}}}{\arxivlink{#2}
  [#3]}}
\providecommand{\eprintlink}[1]{\eprinttmp@#1arXiv: [x]@}
\renewcommand{\eprint}[1]{\eprintlink{#1}}
\providecommand{\adsurl}[1]{\href{#1}{ADS}}
\renewcommand{\bibinfo}[2]{\ifthenelse{\equal{#1}{isbn}}{\href{http://cosmologist.info/ISBN/#2}{#2}}{#2}}
\begin{thebibliography}{53}
\expandafter\ifx\csname natexlab\endcsname\relax\def\natexlab#1{#1}\fi
\expandafter\ifx\csname bibnamefont\endcsname\relax
  \def\bibnamefont#1{#1}\fi
\expandafter\ifx\csname bibfnamefont\endcsname\relax
  \def\bibfnamefont#1{#1}\fi
\expandafter\ifx\csname citenamefont\endcsname\relax
  \def\citenamefont#1{#1}\fi
\expandafter\ifx\csname url\endcsname\relax
  \def\url#1{\texttt{#1}}\fi
\expandafter\ifx\csname urlprefix\endcsname\relax\def\urlprefix{URL }\fi

\bibitem[{\citenamefont{Battye et~al.}(2015)\citenamefont{Battye, Charnock, and
  Moss}}]{Battye:2014qga}
\bibinfo{author}{\bibfnamefont{R.~A.} \bibnamefont{Battye}},
  \bibinfo{author}{\bibfnamefont{T.}~\bibnamefont{Charnock}}, \bibnamefont{and}
  \bibinfo{author}{\bibfnamefont{A.}~\bibnamefont{Moss}},
  \bibinfo{journal}{Phys. Rev.} \textbf{\bibinfo{volume}{D91}},
  \bibinfo{pages}{103508} (\bibinfo{year}{2015}), \eprint{1409.2769}.

\bibitem[{\citenamefont{Ade et~al.}(2015{\natexlab{a}})}]{Ade:2015xua}
\bibinfo{author}{\bibfnamefont{P.~A.~R.} \bibnamefont{Ade}}
  \bibnamefont{et~al.} (\bibinfo{collaboration}{Planck})
  (\bibinfo{year}{2015}{\natexlab{a}}), \eprint{1502.01589}.

\bibitem[{\citenamefont{{Planck Collaboration}
  et~al.}(2013)\citenamefont{{Planck Collaboration}, {Ade}, {Aghanim},
  {Armitage-Caplan}, {Arnaud}, {Ashdown}, {Atrio-Barandela}, {Aumont},
  {Baccigalupi}, {Banday} et~al.}}]{2013arXiv1303.5080P}
\bibinfo{author}{\bibnamefont{{Planck Collaboration}}},
  \bibinfo{author}{\bibfnamefont{P.~A.~R.} \bibnamefont{{Ade}}},
  \bibinfo{author}{\bibfnamefont{N.}~\bibnamefont{{Aghanim}}},
  \bibinfo{author}{\bibfnamefont{C.}~\bibnamefont{{Armitage-Caplan}}},
  \bibnamefont{et~al.}, \bibinfo{journal}{ArXiv e-prints}
  (\bibinfo{year}{2013}), \eprint{1303.5080}.

\bibitem[{\citenamefont{{Riemer-S{\o}rensen}
  et~al.}(2013)\citenamefont{{Riemer-S{\o}rensen}, {Parkinson}, and
  {Davis}}}]{2013arXiv1306.4153R}
\bibinfo{author}{\bibfnamefont{S.}~\bibnamefont{{Riemer-S{\o}rensen}}},
  \bibinfo{author}{\bibfnamefont{D.}~\bibnamefont{{Parkinson}}},
  \bibnamefont{and} \bibinfo{author}{\bibfnamefont{T.~M.}
  \bibnamefont{{Davis}}}, \bibinfo{journal}{ArXiv e-prints}
  (\bibinfo{year}{2013}), \eprint{1306.4153}.

\bibitem[{\citenamefont{Battye and Moss}(2014)}]{Battye:2013}
\bibinfo{author}{\bibfnamefont{R.~A.} \bibnamefont{Battye}} \bibnamefont{and}
  \bibinfo{author}{\bibfnamefont{A.}~\bibnamefont{Moss}},
  \bibinfo{journal}{Phys.Rev.Lett.} \textbf{\bibinfo{volume}{112}},
  \bibinfo{pages}{051303} (\bibinfo{year}{2014}), \eprint{1308.5870}.

\bibitem[{\citenamefont{Gao and Gong}(2014)}]{Gao:2013pfa}
\bibinfo{author}{\bibfnamefont{Q.}~\bibnamefont{Gao}} \bibnamefont{and}
  \bibinfo{author}{\bibfnamefont{Y.}~\bibnamefont{Gong}},
  \bibinfo{journal}{Class. Quant. Grav.} \textbf{\bibinfo{volume}{31}},
  \bibinfo{pages}{105007} (\bibinfo{year}{2014}), \eprint{1308.5627}.

\bibitem[{\citenamefont{Hamann and Hasenkamp}(2013)}]{Hamann:2013}
\bibinfo{author}{\bibfnamefont{J.}~\bibnamefont{Hamann}} \bibnamefont{and}
  \bibinfo{author}{\bibfnamefont{J.}~\bibnamefont{Hasenkamp}},
  \bibinfo{journal}{JCAP} \textbf{\bibinfo{volume}{1310}}, \bibinfo{pages}{044}
  (\bibinfo{year}{2013}), \eprint{1308.3255}.

\bibitem[{\citenamefont{Wyman et~al.}(2014)\citenamefont{Wyman, Rudd,
  Vanderveld, and Hu}}]{Wyman:2013}
\bibinfo{author}{\bibfnamefont{M.}~\bibnamefont{Wyman}},
  \bibinfo{author}{\bibfnamefont{D.~H.} \bibnamefont{Rudd}},
  \bibinfo{author}{\bibfnamefont{R.~A.} \bibnamefont{Vanderveld}},
  \bibnamefont{and} \bibinfo{author}{\bibfnamefont{W.}~\bibnamefont{Hu}},
  \bibinfo{journal}{Phys.Rev.Lett.} \textbf{\bibinfo{volume}{112}},
  \bibinfo{pages}{051302} (\bibinfo{year}{2014}), \eprint{1307.7715}.

\bibitem[{\citenamefont{Beutler et~al.}(2014{\natexlab{a}})}]{Beutler:2014yhv}
\bibinfo{author}{\bibfnamefont{F.}~\bibnamefont{Beutler}} \bibnamefont{et~al.}
  (\bibinfo{collaboration}{BOSS Collaboration})
  (\bibinfo{year}{2014}{\natexlab{a}}), \eprint{1403.4599}.

\bibitem[{\citenamefont{MacCrann et~al.}(2015)\citenamefont{MacCrann, Zuntz,
  Bridle, Jain, and Becker}}]{MacCrann:2014wfa}
\bibinfo{author}{\bibfnamefont{N.}~\bibnamefont{MacCrann}},
  \bibinfo{author}{\bibfnamefont{J.}~\bibnamefont{Zuntz}},
  \bibinfo{author}{\bibfnamefont{S.}~\bibnamefont{Bridle}},
  \bibinfo{author}{\bibfnamefont{B.}~\bibnamefont{Jain}}, \bibnamefont{et~al.},
  \bibinfo{journal}{MNRAS} \textbf{\bibinfo{volume}{451}},
  \bibinfo{pages}{2877} (\bibinfo{year}{2015}), \eprint{1408.4742}.

\bibitem[{\citenamefont{Ruiz and Huterer}(2015)}]{Ruiz:2014hma}
\bibinfo{author}{\bibfnamefont{E.~J.} \bibnamefont{Ruiz}} \bibnamefont{and}
  \bibinfo{author}{\bibfnamefont{D.}~\bibnamefont{Huterer}},
  \bibinfo{journal}{Phys. Rev.} \textbf{\bibinfo{volume}{D91}},
  \bibinfo{pages}{063009} (\bibinfo{year}{2015}), \eprint{1410.5832}.

\bibitem[{\citenamefont{Addison et~al.}(2016)\citenamefont{Addison, Huang,
  Watts, Bennett, Halpern, Hinshaw, and Weiland}}]{Addison:2015wyg}
\bibinfo{author}{\bibfnamefont{G.~E.} \bibnamefont{Addison}},
  \bibinfo{author}{\bibfnamefont{Y.}~\bibnamefont{Huang}},
  \bibinfo{author}{\bibfnamefont{D.~J.} \bibnamefont{Watts}},
  \bibinfo{author}{\bibfnamefont{C.~L.} \bibnamefont{Bennett}},
  \bibnamefont{et~al.}, \bibinfo{journal}{ApJ} \textbf{\bibinfo{volume}{818}},
  \bibinfo{pages}{132} (\bibinfo{year}{2016}), \eprint{1511.00055}.

\bibitem[{\citenamefont{Dossett et~al.}(2015)\citenamefont{Dossett, Ishak,
  Parkinson, and Davis}}]{Dossett:2015nda}
\bibinfo{author}{\bibfnamefont{J.~N.} \bibnamefont{Dossett}},
  \bibinfo{author}{\bibfnamefont{M.}~\bibnamefont{Ishak}},
  \bibinfo{author}{\bibfnamefont{D.}~\bibnamefont{Parkinson}},
  \bibnamefont{and} \bibinfo{author}{\bibfnamefont{T.}~\bibnamefont{Davis}},
  \bibinfo{journal}{Phys. Rev.} \textbf{\bibinfo{volume}{D92}},
  \bibinfo{pages}{023003} (\bibinfo{year}{2015}), \eprint{1501.03119}.

\bibitem[{\citenamefont{Hu and Raveri}(2015)}]{Hu:2015rva}
\bibinfo{author}{\bibfnamefont{B.}~\bibnamefont{Hu}} \bibnamefont{and}
  \bibinfo{author}{\bibfnamefont{M.}~\bibnamefont{Raveri}},
  \bibinfo{journal}{Phys. Rev.} \textbf{\bibinfo{volume}{D91}},
  \bibinfo{pages}{123515} (\bibinfo{year}{2015}), \eprint{1502.06599}.

\bibitem[{\citenamefont{Kitching et~al.}(2016)\citenamefont{Kitching, Verde,
  Heavens, and Jimenez}}]{Kitching:2016hvn}
\bibinfo{author}{\bibfnamefont{T.~D.} \bibnamefont{Kitching}},
  \bibinfo{author}{\bibfnamefont{L.}~\bibnamefont{Verde}},
  \bibinfo{author}{\bibfnamefont{A.~F.} \bibnamefont{Heavens}},
  \bibnamefont{and} \bibinfo{author}{\bibfnamefont{R.}~\bibnamefont{Jimenez}}
  (\bibinfo{year}{2016}), \eprint{1602.02960}.

\bibitem[{\citenamefont{Ade et~al.}(2014{\natexlab{a}})}]{Ade:2013zuv}
\bibinfo{author}{\bibfnamefont{P.~A.~R.} \bibnamefont{Ade}}
  \bibnamefont{et~al.} (\bibinfo{collaboration}{Planck}),
  \bibinfo{journal}{A\&A} \textbf{\bibinfo{volume}{571}}, \bibinfo{pages}{A16}
  (\bibinfo{year}{2014}{\natexlab{a}}), \eprint{1303.5076}.

\bibitem[{\citenamefont{Hinshaw et~al.}(2013)}]{Hinshaw:2012aka}
\bibinfo{author}{\bibfnamefont{G.}~\bibnamefont{Hinshaw}} \bibnamefont{et~al.}
  (\bibinfo{collaboration}{WMAP}), \bibinfo{journal}{ApJS}
  \textbf{\bibinfo{volume}{208}}, \bibinfo{pages}{19} (\bibinfo{year}{2013}),
  \eprint{1212.5226}.

\bibitem[{\citenamefont{Anderson et~al.}(2013)\citenamefont{Anderson, Aubourg,
  Bailey, Bizyaev, Blanton et~al.}}]{Anderson:2012sa}
\bibinfo{author}{\bibfnamefont{L.}~\bibnamefont{Anderson}},
  \bibinfo{author}{\bibfnamefont{E.}~\bibnamefont{Aubourg}},
  \bibinfo{author}{\bibfnamefont{S.}~\bibnamefont{Bailey}},
  \bibinfo{author}{\bibfnamefont{D.}~\bibnamefont{Bizyaev}},
  \bibnamefont{et~al.}, \bibinfo{journal}{Mon.Not.Roy.Astron.Soc.}
  \textbf{\bibinfo{volume}{427}}, \bibinfo{pages}{3435} (\bibinfo{year}{2013}),
  \eprint{1203.6594}.

\bibitem[{\citenamefont{Jones et~al.}(2004)\citenamefont{Jones, Saunders,
  Colless, Read, Parker et~al.}}]{Jones:2004zy}
\bibinfo{author}{\bibfnamefont{D.~H.} \bibnamefont{Jones}},
  \bibinfo{author}{\bibfnamefont{W.}~\bibnamefont{Saunders}},
  \bibinfo{author}{\bibfnamefont{M.}~\bibnamefont{Colless}},
  \bibinfo{author}{\bibfnamefont{M.~A.} \bibnamefont{Read}},
  \bibnamefont{et~al.}, \bibinfo{journal}{Mon.Not.Roy.Astron.Soc.}
  \textbf{\bibinfo{volume}{355}}, \bibinfo{pages}{747} (\bibinfo{year}{2004}),
  \eprint{astro-ph/0403501}.

\bibitem[{\citenamefont{Padmanabhan et~al.}(2012)\citenamefont{Padmanabhan, Xu,
  Eisenstein, Scalzo, Cuesta et~al.}}]{Padmanabhan:2012hf}
\bibinfo{author}{\bibfnamefont{N.}~\bibnamefont{Padmanabhan}},
  \bibinfo{author}{\bibfnamefont{X.}~\bibnamefont{Xu}},
  \bibinfo{author}{\bibfnamefont{D.~J.} \bibnamefont{Eisenstein}},
  \bibinfo{author}{\bibfnamefont{R.}~\bibnamefont{Scalzo}},
  \bibnamefont{et~al.}, \bibinfo{journal}{Mon.Not.Roy.Astron.Soc.}
  \textbf{\bibinfo{volume}{427}}, \bibinfo{pages}{2132} (\bibinfo{year}{2012}),
  \eprint{1202.0090}.

\bibitem[{\citenamefont{Heymans et~al.}(2013)}]{Heymans:2013fya}
\bibinfo{author}{\bibfnamefont{C.}~\bibnamefont{Heymans}} \bibnamefont{et~al.},
  \bibinfo{journal}{MNRAS} \textbf{\bibinfo{volume}{432}},
  \bibinfo{pages}{2433} (\bibinfo{year}{2013}), \eprint{1303.1808}.

\bibitem[{\citenamefont{Ade et~al.}(2014{\natexlab{b}})}]{Ade:2013tyw}
\bibinfo{author}{\bibfnamefont{P.~A.~R.} \bibnamefont{Ade}}
  \bibnamefont{et~al.} (\bibinfo{collaboration}{Planck}),
  \bibinfo{journal}{A\&A} \textbf{\bibinfo{volume}{571}}, \bibinfo{pages}{A17}
  (\bibinfo{year}{2014}{\natexlab{b}}), \eprint{1303.5077}.

\bibitem[{\citenamefont{Beutler et~al.}(2014{\natexlab{b}})}]{Beutler:2013yhm}
\bibinfo{author}{\bibfnamefont{F.}~\bibnamefont{Beutler}} \bibnamefont{et~al.}
  (\bibinfo{collaboration}{BOSS}), \bibinfo{journal}{MNRAS}
  \textbf{\bibinfo{volume}{443}}, \bibinfo{pages}{1065}
  (\bibinfo{year}{2014}{\natexlab{b}}), \eprint{1312.4611}.

\bibitem[{\citenamefont{Lewis and Bridle}(2002)}]{Lewis:2002ah}
\bibinfo{author}{\bibfnamefont{A.}~\bibnamefont{Lewis}} \bibnamefont{and}
  \bibinfo{author}{\bibfnamefont{S.}~\bibnamefont{Bridle}},
  \bibinfo{journal}{Phys. Rev.} \textbf{\bibinfo{volume}{D66}},
  \bibinfo{pages}{103511} (\bibinfo{year}{2002}), \eprint{astro-ph/0205436}.

\bibitem[{\citenamefont{Inman and Jr}(1989)}]{Inman}
\bibinfo{author}{\bibfnamefont{H.~F.} \bibnamefont{Inman}} \bibnamefont{and}
  \bibinfo{author}{\bibfnamefont{E.~L.~B.} \bibnamefont{Jr}},
  \bibinfo{journal}{Communications in Statistics - Theory and Methods}
  \textbf{\bibinfo{volume}{18}}, \bibinfo{pages}{3851} (\bibinfo{year}{1989}),
  \eprint{http://dx.doi.org/10.1080/03610928908830127},
  \urlprefix\url{http://dx.doi.org/10.1080/03610928908830127}.

\bibitem[{\citenamefont{Bhattacharyya}(1943)}]{Bhattacharyya}
\bibinfo{author}{\bibfnamefont{A.}~\bibnamefont{Bhattacharyya}},
  \bibinfo{journal}{Calcutta Math. Soc. Bull.} \textbf{\bibinfo{volume}{35}},
  \bibinfo{pages}{99} (\bibinfo{year}{1943}).

\bibitem[{\citenamefont{Marshall et~al.}(2006)\citenamefont{Marshall, Rajguru,
  and Slosar}}]{Marshall:2004zd}
\bibinfo{author}{\bibfnamefont{P.}~\bibnamefont{Marshall}},
  \bibinfo{author}{\bibfnamefont{N.}~\bibnamefont{Rajguru}}, \bibnamefont{and}
  \bibinfo{author}{\bibfnamefont{A.}~\bibnamefont{Slosar}},
  \bibinfo{journal}{Phys. Rev.} \textbf{\bibinfo{volume}{D73}},
  \bibinfo{pages}{067302} (\bibinfo{year}{2006}), \eprint{astro-ph/0412535}.

\bibitem[{\citenamefont{Verde et~al.}(2013)\citenamefont{Verde, Protopapas, and
  Jimenez}}]{Verde:2013wza}
\bibinfo{author}{\bibfnamefont{L.}~\bibnamefont{Verde}},
  \bibinfo{author}{\bibfnamefont{P.}~\bibnamefont{Protopapas}},
  \bibnamefont{and} \bibinfo{author}{\bibfnamefont{R.}~\bibnamefont{Jimenez}},
  \bibinfo{journal}{Phys. Dark Univ.} \textbf{\bibinfo{volume}{2}},
  \bibinfo{pages}{166} (\bibinfo{year}{2013}), \eprint{1306.6766}.

\bibitem[{\citenamefont{Seehars et~al.}(2014)\citenamefont{Seehars, Amara,
  Refregier, Paranjape, and Akeret}}]{Seehars:2014ora}
\bibinfo{author}{\bibfnamefont{S.}~\bibnamefont{Seehars}},
  \bibinfo{author}{\bibfnamefont{A.}~\bibnamefont{Amara}},
  \bibinfo{author}{\bibfnamefont{A.}~\bibnamefont{Refregier}},
  \bibinfo{author}{\bibfnamefont{A.}~\bibnamefont{Paranjape}},
  \bibnamefont{et~al.}, \bibinfo{journal}{Phys. Rev.}
  \textbf{\bibinfo{volume}{D90}}, \bibinfo{pages}{023533}
  (\bibinfo{year}{2014}), \eprint{1402.3593}.

\bibitem[{\citenamefont{Rice}(2006)}]{Rice:2006}
\bibinfo{author}{\bibfnamefont{J.}~\bibnamefont{Rice}},
  \emph{\bibinfo{title}{Mathematical Statistics and Data Analysis}}, Advanced
  series (\bibinfo{publisher}{Cengage Learning}, \bibinfo{year}{2006}).

\bibitem[{\citenamefont{Rudary}(2009)}]{rudary2009predictive}
\bibinfo{author}{\bibfnamefont{M.~R.} \bibnamefont{Rudary}},
  \emph{\bibinfo{title}{On Predictive Linear Gaussian Models}}
  (\bibinfo{publisher}{ProQuest}, \bibinfo{year}{2009}).

\bibitem[{\citenamefont{Adam et~al.}(2015)}]{Adam:2015rua}
\bibinfo{author}{\bibfnamefont{R.}~\bibnamefont{Adam}} \bibnamefont{et~al.}
  (\bibinfo{collaboration}{Planck}) (\bibinfo{year}{2015}),
  \eprint{1502.01582}.

\bibitem[{\citenamefont{Aghanim et~al.}(2015)}]{Aghanim:2015xee}
\bibinfo{author}{\bibfnamefont{N.}~\bibnamefont{Aghanim}} \bibnamefont{et~al.}
  (\bibinfo{collaboration}{Planck}), \bibinfo{journal}{Submitted to: Astron.
  Astrophys.}  (\bibinfo{year}{2015}), \eprint{1507.02704}.

\bibitem[{\citenamefont{Beutler et~al.}(2011)\citenamefont{Beutler, Blake,
  Colless, Jones, Staveley-Smith et~al.}}]{Beutler:2011hx}
\bibinfo{author}{\bibfnamefont{F.}~\bibnamefont{Beutler}},
  \bibinfo{author}{\bibfnamefont{C.}~\bibnamefont{Blake}},
  \bibinfo{author}{\bibfnamefont{M.}~\bibnamefont{Colless}},
  \bibinfo{author}{\bibfnamefont{D.~H.} \bibnamefont{Jones}},
  \bibnamefont{et~al.}, \bibinfo{journal}{Mon.Not.Roy.Astron.Soc.}
  \textbf{\bibinfo{volume}{416}}, \bibinfo{pages}{3017} (\bibinfo{year}{2011}),
  \eprint{1106.3366}.

\bibitem[{\citenamefont{Gil-Marín
  et~al.}(2015{\natexlab{a}})}]{Gil-Marin:2015nqa}
\bibinfo{author}{\bibfnamefont{H.}~\bibnamefont{Gil-Marín}}
  \bibnamefont{et~al.} (\bibinfo{year}{2015}{\natexlab{a}}),
  \eprint{1509.06373}.

\bibitem[{\citenamefont{Joudaki et~al.}(2016)}]{Joudaki:2016mvz}
\bibinfo{author}{\bibfnamefont{S.}~\bibnamefont{Joudaki}} \bibnamefont{et~al.}
  (\bibinfo{year}{2016}), \eprint{1601.05786}.

\bibitem[{\citenamefont{Kaiser et~al.}(2000)\citenamefont{Kaiser, Wilson, and
  Luppino}}]{Kaiser:2000if}
\bibinfo{author}{\bibfnamefont{N.}~\bibnamefont{Kaiser}},
  \bibinfo{author}{\bibfnamefont{G.}~\bibnamefont{Wilson}}, \bibnamefont{and}
  \bibinfo{author}{\bibfnamefont{G.~A.} \bibnamefont{Luppino}}
  (\bibinfo{year}{2000}), \eprint{astro-ph/0003338}.

\bibitem[{\citenamefont{Abbott et~al.}(2015)}]{Abbott:2015swa}
\bibinfo{author}{\bibfnamefont{T.}~\bibnamefont{Abbott}} \bibnamefont{et~al.}
  (\bibinfo{collaboration}{DES}) (\bibinfo{year}{2015}), \eprint{1507.05552}.

\bibitem[{\citenamefont{Kuijken et~al.}(2015)}]{Kuijken:2015vca}
\bibinfo{author}{\bibfnamefont{K.}~\bibnamefont{Kuijken}} \bibnamefont{et~al.},
  \bibinfo{journal}{MNRAS} \textbf{\bibinfo{volume}{454}},
  \bibinfo{pages}{3500} (\bibinfo{year}{2015}), \eprint{1507.00738}.

\bibitem[{\citenamefont{Lewis and Challinor}(2006)}]{Lewis:2006fu}
\bibinfo{author}{\bibfnamefont{A.}~\bibnamefont{Lewis}} \bibnamefont{and}
  \bibinfo{author}{\bibfnamefont{A.}~\bibnamefont{Challinor}},
  \bibinfo{journal}{Phys. Rept.} \textbf{\bibinfo{volume}{429}},
  \bibinfo{pages}{1} (\bibinfo{year}{2006}), \eprint{astro-ph/0601594}.

\bibitem[{\citenamefont{Ade et~al.}(2015{\natexlab{b}})}]{Ade:2015zua}
\bibinfo{author}{\bibfnamefont{P.~A.~R.} \bibnamefont{Ade}}
  \bibnamefont{et~al.} (\bibinfo{collaboration}{Planck})
  (\bibinfo{year}{2015}{\natexlab{b}}), \eprint{1502.01591}.

\bibitem[{\citenamefont{Gil-Marín
  et~al.}(2015{\natexlab{b}})}]{Gil-Marin:2015sqa}
\bibinfo{author}{\bibfnamefont{H.}~\bibnamefont{Gil-Marín}}
  \bibnamefont{et~al.} (\bibinfo{year}{2015}{\natexlab{b}}),
  \eprint{1509.06386}.

\bibitem[{\citenamefont{Ade et~al.}(2015{\natexlab{c}})}]{Ade:2015fva}
\bibinfo{author}{\bibfnamefont{P.~A.~R.} \bibnamefont{Ade}}
  \bibnamefont{et~al.} (\bibinfo{collaboration}{Planck})
  (\bibinfo{year}{2015}{\natexlab{c}}), \eprint{1502.01597}.

\bibitem[{\citenamefont{Mantz et~al.}(2015)}]{Mantz:2014paa}
\bibinfo{author}{\bibfnamefont{A.~B.} \bibnamefont{Mantz}}
  \bibnamefont{et~al.}, \bibinfo{journal}{MNRAS}
  \textbf{\bibinfo{volume}{446}}, \bibinfo{pages}{2205} (\bibinfo{year}{2015}),
  \eprint{1407.4516}.

\bibitem[{\citenamefont{Jeffreys}(1961)}]{Jeffreys:1961}
\bibinfo{author}{\bibfnamefont{S.~H.} \bibnamefont{Jeffreys}},
  \emph{\bibinfo{title}{The Theory of Probability}} (\bibinfo{publisher}{Oxford
  University Press}, \bibinfo{address}{New York, NY, USA},
  \bibinfo{year}{1961}).

\bibitem[{\citenamefont{Bond et~al.}(1980)\citenamefont{Bond, Efstathiou, and
  Silk}}]{Bond:1980}
\bibinfo{author}{\bibfnamefont{J.}~\bibnamefont{Bond}},
  \bibinfo{author}{\bibfnamefont{G.}~\bibnamefont{Efstathiou}},
  \bibnamefont{and} \bibinfo{author}{\bibfnamefont{J.}~\bibnamefont{Silk}},
  \bibinfo{journal}{Phys.Rev.Lett.} \textbf{\bibinfo{volume}{45}},
  \bibinfo{pages}{1980} (\bibinfo{year}{1980}).

\bibitem[{\citenamefont{Hu et~al.}(1998)\citenamefont{Hu, Eisenstein, and
  Tegmark}}]{Hu:1997}
\bibinfo{author}{\bibfnamefont{W.}~\bibnamefont{Hu}},
  \bibinfo{author}{\bibfnamefont{D.~J.} \bibnamefont{Eisenstein}},
  \bibnamefont{and} \bibinfo{author}{\bibfnamefont{M.}~\bibnamefont{Tegmark}},
  \bibinfo{journal}{Phys.Rev.Lett.} \textbf{\bibinfo{volume}{80}},
  \bibinfo{pages}{5255} (\bibinfo{year}{1998}), \eprint{astro-ph/9712057}.

\bibitem[{\citenamefont{Elgaroy and Lahav}(2005)}]{Elgaroy:2004}
\bibinfo{author}{\bibfnamefont{O.}~\bibnamefont{Elgaroy}} \bibnamefont{and}
  \bibinfo{author}{\bibfnamefont{O.}~\bibnamefont{Lahav}},
  \bibinfo{journal}{New J.Phys.} \textbf{\bibinfo{volume}{7}},
  \bibinfo{pages}{61} (\bibinfo{year}{2005}), \eprint{hep-ph/0412075}.

\bibitem[{\citenamefont{Lesgourgues and Pastor}(2006)}]{Lesgourgues:2006}
\bibinfo{author}{\bibfnamefont{J.}~\bibnamefont{Lesgourgues}} \bibnamefont{and}
  \bibinfo{author}{\bibfnamefont{S.}~\bibnamefont{Pastor}},
  \bibinfo{journal}{Phys.Rept.} \textbf{\bibinfo{volume}{429}},
  \bibinfo{pages}{307} (\bibinfo{year}{2006}), \eprint{astro-ph/0603494}.

\bibitem[{\citenamefont{Hannestad}(2010)}]{Hannestad:2010}
\bibinfo{author}{\bibfnamefont{S.}~\bibnamefont{Hannestad}},
  \bibinfo{journal}{Prog.Part.Nucl.Phys.} \textbf{\bibinfo{volume}{65}},
  \bibinfo{pages}{185} (\bibinfo{year}{2010}), \eprint{1007.0658}.

\bibitem[{\citenamefont{Hall and Challinor}(2012)}]{Hall:2012}
\bibinfo{author}{\bibfnamefont{A.~C.} \bibnamefont{Hall}} \bibnamefont{and}
  \bibinfo{author}{\bibfnamefont{A.}~\bibnamefont{Challinor}},
  \bibinfo{journal}{Mon.Not.Roy.Astron.Soc.} \textbf{\bibinfo{volume}{425}},
  \bibinfo{pages}{1170} (\bibinfo{year}{2012}), \eprint{1205.6172}.

\bibitem[{\citenamefont{Leistedt et~al.}(2014)\citenamefont{Leistedt, Peiris,
  and Verde}}]{Boris}
\bibinfo{author}{\bibfnamefont{B.}~\bibnamefont{Leistedt}},
  \bibinfo{author}{\bibfnamefont{H.~V.} \bibnamefont{Peiris}},
  \bibnamefont{and} \bibinfo{author}{\bibfnamefont{L.}~\bibnamefont{Verde}}
  (\bibinfo{year}{2014}), \eprint{1404.5950}.

\bibitem[{\citenamefont{Adam et~al.}(2016)}]{Adam:2016hgk}
\bibinfo{author}{\bibfnamefont{R.}~\bibnamefont{Adam}} \bibnamefont{et~al.}
  (\bibinfo{collaboration}{Planck}) (\bibinfo{year}{2016}),
  \eprint{1605.03507}.

\end{thebibliography}

    \begin{appendices}
\section{Comparison of methods\label{Appendix}}
            To understand how each of the different methods of quantification work it is useful to compare some simple distributions, shown in tables~\ref{t:1Ddistributions} and~\ref{t:2Ddistributions}.
            Figures for each measure of every 1D and 2D parameter distribution comparison can be found on pages~\pageref{fig:I}-\pageref{fig:V}.
            As in Sec.~\ref{S:Tension}, these posterior distributions are $P_1\equiv P(\btheta|D_1,\mathcal{M})$ and $P_2\equiv P(\btheta|D_2,\mathcal{M})$ for data sets $D_1$ and $D_2$ respectively in a model $\mathcal{M}$.
            In the 1D case $\btheta\equiv\theta$ is a one dimensional parameter, whereas in 2D $\btheta=\{\theta_1,\theta_2\}$.
            In each case the probability distributions are normalised such that%
            \begin{equation}
                \int d\btheta P_i = 1.
            \end{equation}
            \begin{table}
                \centering
                \begin{tabular}{lll}\hline\hline
                    & $P_1\hskip10em$ & $P_2\hskip10em$\\\hline
                    I & $\mathcal{N}(0,1)$ & $\mathcal{N}(0,1)$ \\
                    II & $\mathcal{N}(0,1)$ & $\mathcal{N}(0,3)$ \\
                    III & $\mathcal{N}(5,1)$ & $\mathcal{N}(-5,1)$ \\
                    IV & $\mathcal{N}(0,1)$ & $\mathcal{N}(1.427, 1)$ \\
                    V & $\mathcal{N}(0,1)+\mathcal{N}(-2,1)$ & $\mathcal{N}(1.427,1)+\mathcal{N}(4,2)$\\\hline\hline
                \end{tabular}
                \caption[1D probability distributions being compared]{1D probability distributions being compared.}
                \label{t:1Ddistributions}
            \end{table}
            \begin{table}
                \centering
                \begin{tabular}{lll}\hline\hline
                    & $P_1$ & $P_2$ \\\hline
                    I & $\mathcal{N}\overset{~}{\left(\begin{pmatrix}0&0\end{pmatrix},\begin{pmatrix}1&0\\0&1\end{pmatrix}\right)}$ & $\mathcal{N}\left(\begin{pmatrix}0&0\end{pmatrix},\begin{pmatrix}1&0\\0&1\end{pmatrix}\right)$ \\
                    II & $\mathcal{N}\overset{~}{\left(\begin{pmatrix}0&0\end{pmatrix},\begin{pmatrix}1&0\\0&1\end{pmatrix}\right)}$ & $\mathcal{N}\left(\begin{pmatrix}0&0\end{pmatrix},\begin{pmatrix}3^2&0\\0&3^2\end{pmatrix}\right)$ \\
                    III & $\mathcal{N}\overset{~}{\left(\begin{pmatrix}5&5\end{pmatrix},\begin{pmatrix}1&0\\0&1\end{pmatrix}\right)}$ & $\mathcal{N}\left(\begin{pmatrix}-5&-5\end{pmatrix},\begin{pmatrix}1&0\\0&1\end{pmatrix}\right)$ \\
                    IV & $\mathcal{N}\overset{~}{\underset{~}{\left(\begin{pmatrix}0&0\end{pmatrix},\begin{pmatrix}1&0\\0&1\end{pmatrix}\right)}}$& $\mathcal{N}\left(\begin{pmatrix}1.427&1.427\end{pmatrix},\begin{pmatrix}1&0\\0&1\end{pmatrix}\right)$ \\
                    V & $\mathcal{N}\overset{~}{\left(\begin{pmatrix}0&0\end{pmatrix},\begin{pmatrix}1&0\\0&1\end{pmatrix}\right)}$~~~& $\mathcal{N}\left(\begin{pmatrix}1.427&1.427\end{pmatrix},\begin{pmatrix}1&0\\0&1\end{pmatrix}\right)$ \\
                    &\multicolumn{2}{l}{$+\mathcal{N}\overset{~}{\underset{~}{\left(\begin{pmatrix}-2&-2\end{pmatrix},\begin{pmatrix}1&0\\0&1\end{pmatrix}\right)}}$~~~$+\mathcal{N}\overset{~}{\left(\begin{pmatrix}4&4\end{pmatrix},\begin{pmatrix}2^2&0\\0&2^2\end{pmatrix}\right)}$}\\\hline\hline
                \end{tabular}
                \caption[2D probability distributions being compared]{2D probability distributions being compared.}
                \label{t:2Ddistributions}
            \end{table}
            \paragraph*{I - Identical distributions} Figure~\ref{fig:I} shows the distributions and integrated measures quantifying the amount of agreement or disagreement of two identical distributions, described in row I of tables~\ref{t:1Ddistributions} and~\ref{t:2Ddistributions}.
            Each method is unanimous in its quantification of the combination of these two distributions in both 1D and 2D.
            \\
            \indent\emph{1, 2.} The Bhattacharyya distance and the overlap coefficient are $B=1$ and $O=1$, in both one and two dimensions.
            This shows the distributions are identical, since $P_1=P_2$ then $\sqrt{P_1P_2}=P_1=P_2$ and ${\rm Min}[P_1, P_2] = P_1 = P_2$ which is unity when integrated as in equations~(\ref{E:B}) and~(\ref{E:O}).\\
            \indent\emph{3.} A value of $C=0$ means that the distributions must be identical.
            The parameter ranges are identical for identical distributions (and infinite for the distributions in tables~\ref{t:1Ddistributions} and~\ref{t:2Ddistributions}) so the difference in the range is the same, $\delta\theta=\theta_1=\theta_2$.
            A new Gaussian is formed with half the variance and a mean at $\delta\theta=0$.
            Since $\dtheta=0$ is at the maximum of the distribution then there are no parameter ranges above the value of the probability distribution function at $\delta=0$ to integrate.
            For the result in figure~\ref{fig:I} the result obtained by integrating inside the isocontour formed by the value of the probability density function at $\delta\theta=0$ deviates slightly from zero due to the finite number of samples taken.
            The rest of the samples outside of this boundary can be considered consistent.\\
            \indent\emph{4.} When $I_1=I_2=0.997$ then the two distributions are shown to be identical.
            The set of parameter values which contain 99.7\% of the samples of either distribution are equal for identical distributions.
            This means integrating either distribution for these parameter ranges will equate to $I_{i}=0.997$, i.e. the total fraction of samples that can be drawn from the parameter ranges are drawn from both distributions. \\
            \indent\emph{5.} There is no gain in information when two distributions are identical.
            Since this is expected it means there is also no ``surprise''.
            This can be seen trivially in equation~(\ref{E:RelativeEntropy}) since $\log(P_2/P_1)=\log1=0$ when $P_1=P_2$.\\
            \indent\emph{6.} A value of $\log R=1.730$ or $\log R=3.460$ shows that evidence favours the combined probability distribution when the distribution is chosen to be uniform between $-10<\theta<10$ in one or two dimensions.
            This is expected for identical distributions since, although the integrals $p(D_1)$ and $p(D_2)$ are greater than $p(D_1, D_2)$ their combination $p(D_1)p(D_2)$ is smaller.
            This is always true independent of the choice of prior. 
            The magnitude of $\log R$ does depend on the prior: $\log R$ is larger when the prior is wider; it is smaller when the prior is narrower.
            The positive $\log R$ can be interpreted as an indication that the two distributions are somewhat similar.
            Although $\log R=1.730$ \emph{means} the distributions are identical with the given prior, it is not a particularly intuitive value.\\
            \indent\emph{7.} Similar to measure~\emph{3}, $\log T=0$ shows that the two distributions are identical since $p(D_1, D_2)_{\rm shifted}=p(D_1, D_2)$.
            Both of the means of the joint probability distributions are the same so the mean of the shifted distribution does not move.
            The ratio is therefore $T=1$ giving a $\log T=0$ showing that they are identical.
            This is again true in both 1D and 2D.
            \paragraph*{II - One distribution broader than the other but with the same mean} Figure~\ref{fig:II} shows the measure of discordance when one distribution remains the same as in I, but the width of the second distribution increases to $\sigma = 3$ as in the second row of tables~\ref{t:1Ddistributions} and~\ref{t:2Ddistributions}.
            A useful measure here would indicate either that the distributions are very similar, or that $P_1$ is completely consistent with $P_2$ even though $P_2$ is not completely consistent with $P_1$.
            \\
            \indent\emph{1.} $B=0.775$ and $B=0.600$ in one and two dimensions.
            These values show that the distributions are not concordant in some way.
            It does not illuminate in which way the distributions disagree.
            Knowing the distributions, it can be seen that the disagreement occurs because the value of $P_2$ are small for parameter values where $P_1$ is large, and vice versa.
            The integral over the combined distributions is therefore less than unity.
            \\
            \indent\emph{2.} Similarly, the overlap coefficient reveals $O=0.516$ and $O=0.325$ in one and two dimensions respectively.
            The low maximum value of $P_2$ means that ${\rm Min}[P_1, P_2]$ is capped where $P_1$ is large.
            This gives the same misleading interpretation as the Bhattacharyya distance.
            In fact, since the values of $O$ are lower, they could be interpreted as the distributions being in greater disagreement.
            \\
            \indent\emph{3.} The measure here does not take into account broadening of distributions and so $C=0$ again.
            The variance of $P_2$ has increased (compared to in I) so the variance of the new distribution $P(\dtheta)$ is larger, but the mean is still centred on $\dtheta=0$.
            The isocontour defined by the value of $P(0)$ contains no parameter values and so integrating again gives zero.
            This measure indicates that the samples in the new distribution are \emph{consistent} and so the original distributions agree.
            In fact, they can be interpreted as being identical, which may be misleading.
            \\
            \indent\emph{4.} This measure is the most informative of all the quantifications of the level of agreement.
            $I_1=1.000$ and $I_2=0.684$ show that all of samples drawn from $P_1$ are contained in the parameter ranges which contain 99.7\% of the samples drawn from $P_2$.
            Simply, $P_1$ is completely consistent with $P_2$.
            $I_2<I_1$ indicates that $P_2$ has a greater variance than $P_1$, the value of $I_2$ showing how broad the distribution is in comparison to $P_1$.
            If $I_2\lesssim I_1$ then $P_2$ is quite similar to $P_1$, but if $I_2\ll I_1$ then $P_2$ has a much greater variance.
            \\
            \indent\emph{5.} There is a gain in information from updating $P_1$ with $P_2$ since there is an extension of available parameter space, but this is mostly due to ``surprise'' as the entropy expected by broadening the distribution is small.
            On the other hand, when $P_1$ updates $P_2$ there is a much smaller relative entropy, but there is expected to be a large amount, so the ``surprise'' is negative.
            These two values can be interpreted as showing that $P_2$ does not agree with $P_1$ as much as expected and that $P_1$ agrees with $P_2$ \emph{more} than is expected.
            \\
            \indent\emph{6.} The interpretation of this measure is exactly the same as for I.
            The distributions must be similar since $\log R$ is positive.
            The value is lower for the same reason that the Bhattacharyya distance is less but, because it is normalised by the evidences of each distribution, it is still informative.
            As such it is possible to tell that, for a given prior, $P_1$ is not the same as $P_2$, but they are still similar. 
            \\
            \indent\emph{7.} Similar to measure~\emph{3}, $\log T=0$ shows the distributions are consistent (or identical in fact). The maximum value of the distribution $p(D_1, D_2)_{\rm shifted}$ is less, but it is still equal to $p(D_1, D_2)$ and so the $\log$ of their ratio vanishes.
            \paragraph*{III - Discordant distributions} Figure~\ref{fig:III} shows examples of each of the measures when two distributions are greatly separated.
            This is the last of the distribution combinations in which all of the measures are in agreement, showing that the distributions are not similar.
            \\
            \indent\emph{1, 2, 4, 6.} Since $P_1$ is negligible where $P_2\ne0$ then the integration of any combination of $P_1$ and $P_2$ will (approximately) vanish, which explains the values of $B=0$ and $O=0$.
            Similarly, if the integration ranges where 99.7\% of the samples from one distribution would be drawn do not overlap with the non-negligible regions of the other distribution then $I_{i}\approx0$.
            Since $p(D_1)$ and $p(D_2)$ are much greater than $P(D_1,D_2)$ (which almost vanishes) then $\log R$ is extremely negative, preferring either evidence to the joint evidence.
            All these measures show that $P_1$ is not at all similar to $P_2$.
            \\
            \indent\emph{3.} The mean of the new distribution is far $\dtheta=0$ and the value of the distribution is negligible there.
            The parameter range within the contour formed where $P(0)=0$ contains the whole distribution and as such $C=1$.
            This is only possible when the whole distribution is integrated, showing that none of the samples drawn from either of the original distributions would be consistent with the other.
            \\
            \indent\emph{5.} There is a very large relative entropy since the distributions contain completely different areas of parameter space, so a large amount of information is gained.
            However, since the means are incompatible, this information is not expected so the whole of the relative entropy is driven by ``surprise''.
            This shows that the distributions do not agree with each other.
            \\
            \indent\emph{7.} When means of $P_2$ are shifted to coincide with the means of $P_1$, $p(D_1,D_2)_{\rm shifted}\gg p(D_1,D_2)$ and so $T$ is large.
            A large positive $\log T$ indicates that the distributions are severely discordant.
            \paragraph*{IV - Slightly shifted distribution} Figure~\ref{fig:IV} shows the row~IV distributions from tables~\ref{t:1Ddistributions} and~\ref{t:2Ddistributions}.
            The second distribution $P_2$, has the same variance as $P_1$ but the means of $P_2$ are shifted such that the value of $B$ is the same as using the distributions in row~II of tables~\ref{t:1Ddistributions} and ~\ref{t:2Ddistributions}.
            \\
            \indent\emph{1.} As already described, $B=0.775$ and $B=0.601$ in one and two dimensions.
            These are the same values obtained when the variance of $P_2$ is three times that of $P_1$.
            This example shows how the Bhattacharyya distance allows broadening of distributions to be mapped to shifts in the mean.
            Due to this, it is harder to interpret the meaning of $B$ without seeing at least a projection of the probability distribution.
            $0<B<1$ could arise from purely a flattening of a distribution, or a shift in the means, or a combination of both.
            \\
            \indent\emph{2.} The overlap coefficient is similar to the Bhattacharyya distance, although a shift in the means of one distribution is more heavily penalised (a lower value of $O$ found) than a broadening of the variance of that distribution.
            The same problem still persists, that there is no distinction between flattening of the distribution or shifts or combinations of them both.\\
            \indent\emph{3.} $P(\dtheta)$ is centred slightly away from $\dtheta=0$ because the means of $P_1$ and $P_2$ are not equal.
            The value of the probability distribution at $\dtheta=0$ forms a contour (or interval) which contains 69.7\% and 61.1\% of the samples drawn from the distribution in one and two dimensions.
            These percentages can be mapped to the proportion of samples drawn from a one dimensional Gaussian, comparing the intervals to a number of standard deviations.
            In this paper, 69.7\% would map to a tension of $\sim1.0\sigma$, which means the distributions are consistent.
            \\
            \indent\emph{4.} Since $I_1=I_2=0.942$ then both $P_1$ and $P_2$ must have the same variance, but $I_1=I_2<0.997$ shows that not all the possible samples are contained within the integration interval.
            This indicates that the means of $P_1$ must not coincide with the means of $P_2$.
            Since the result of $I_1$ and $I_2$ are close to 0.997, then the means are not well separated and hence the distributions are in reasonable agreement.
            \\
            \indent\emph{5.} The information gain from updating either distribution with the other is equal showing that both distributions have the same variance.
            In one dimension this gain is mostly expected and so the ``surprise'' is small and the distributions can be considered compatible.
            In two dimensions there is a lot more expected relative entropy than information gained and so the ``surprise'' is highly negative.
            This means the distributions are more similar \emph{than expected}.
            It is difficult to quantify what this means in terms of similarity of the two distributions.
            \\
            \indent\emph{6.} $\log R=1.221$ and $\log R=2.442$ in one and two dimensions.
            These values indicate that the joint evidence is more likely than each of the individual evidences $p(D_1)$ and $p(D_2)$, and therefore the distributions are similar.
            Interestingly, these measures show that the shift in the means of one of the distributions is \emph{more} consistent than each of the distributions having equal means, but the variance of one being larger (as in row~II of tables~\ref{t:1Ddistributions} and~\ref{t:2Ddistributions}).
            \\
            \indent\emph{7.} $\log T=0.509$ and $\log T=1.018$ show that the distributions are similar but not identical, in one and two dimensions.
            The shifted joint evidence is slightly larger than $p(D_1,D_2)$, but because the means of $P_2$ are close to the means of $P_1$ the ratio between $p(D_1,D_2)_{\rm shifted}$ and $p(D_1,D_2)$ is only slightly greater than one.
            \paragraph*{V - Unusually shaped distributions} Figure~\ref{fig:V} shows the values each of the measures give for unusual shaped distributions (constructed by combining Gaussians in this case) in tables~\ref{t:1Ddistributions} and~\ref{t:2Ddistributions}.
            \\
            \indent\emph{1, 2.} The Bhattacharyya distance is lower than the comparisons of $P_1$ and $P_2$ in rows~II and~IV from tables~\ref{t:1Ddistributions} and \ref{t:2Ddistributions} suggesting that these distributions agree less than in those cases.
            The same is true for the overlap coefficient.
            In the one dimensional case, mapping $B=0.487$ to a shift in the mean only is equivalent to moving the peak of a Gaussian distribution by $\theta=2.4$ from the centre of the other distribution.
            Likewise, $O=0.264$ obtained here is equivalent to shifting the peak of a Gaussian distribution to $\theta=2.2$ compared to another Gaussian with the same variance centred at $\theta=0$.
            Comparing the values of $B$ and $O$ to shifts in the mean is a useful way to interpret results from these methods, although it still does not take into account the flattening of the distributions.
            \\
            \indent\emph{3.} The values of $C=0.620$ and $C=0.310$ suggest that $P_1$ and $P_2$ are extremely consistent, although not identical.
            Mapping to one dimensional Gaussian distributions, these are equivalent to tensions of $0.9\sigma$ and $0.4\sigma$ respectively.
            This maybe quite misleading since (according to figure~\ref{fig:V}) a lot of the distribution lies away from $\dtheta=0$, it is just the primary peak which is near to $\dtheta=0$.
            This means it is the only measure here to quantify these two distributions as more consistent than in row~IV of tables~\ref{t:1Ddistributions} and~\ref{t:2Ddistributions}.
            \\
            \indent\emph{4.} 70.4\% of samples drawn from $P_1$ are within the the 99.7\% confidence intervals of $P_2$ and 59.3\% of the samples drawn from $P_2$ are within the 99.7\% confidence intervals of $P_1$ in one dimension.
            This shows that $P_1$ is more consistent with $P_2$ than the other way around.
            Since $I_1>I_2$ then the effective variance of $P_2$ is larger than $P_1$'s. 
            Both the values of $I_1$ and $I_2$ being less than 0.997 suggests a shift so that the peaks of the distribution are not aligned.
            Of course, the distributions could both be peaked at the same parameter value but one of the distributions skewed which would give similar results.
            \\
            \indent\emph{5.} The relative entropy is mostly ``surprise'' driven suggesting the distributions are not in a great level of agreement.
             $P_2$ is less consistent with $P_1$ than $P_1$ is with $P_2$ since the information gain and ``surprise' are smaller when $P_1$ is used to update $P_2$.
            \\
            \indent\emph{6.} The positive values of $\log R=0.120$ and $\log R=1.110$ show that the two probability distributions are consistent since the joint evidence is more likely than either of the evidences combined.
            The values of $\log R$ are closer to zero than any of the previous comparisons from tables~\ref{t:1Ddistributions} and~\ref{t:2Ddistributions} with the exception of row~III suggesting that the agreement is less in this case.
            \\
            \indent\emph{7.} The ratio of $p(D_1,D_2)_{\rm shifted}$ to $p(D_1,D_2)$ is fairly large so $\log T$ shows that the agreement is less than for the other comparisons in tables~\ref{t:1Ddistributions} and~\ref{t:2Ddistributions} except row~III.
            The value is much less than $\log T$ for row~III and so it is clear that these distributions are not wholly discordant.
            \\
            \\
            When comparing the one and two dimensional distributions it can be seen that the general trends are the same.
            It should be noted here that the 2D distributions are slightly more distinct than the 1D distributions are for each row in tables~\ref{t:1Ddistributions} and~\ref{t:2Ddistributions} so the measure values are expected to show less consistency.
            The integration between interval (\emph{4}) and difference vector methods (\emph{3}) have the same interpretation value independent of the number of dimensions.
            The other methods (\emph{1}, \emph{2}, \emph{5}, \emph{6} and \emph{7}) give different values in different dimensions, which needs to be taken into account or corrected when analysing the measures.
        \subsection{Probability distribution comparison figures\label{App:Figs}}
        
\begin{figure*}
                     \centering
                     \includegraphics[width=\textwidth]{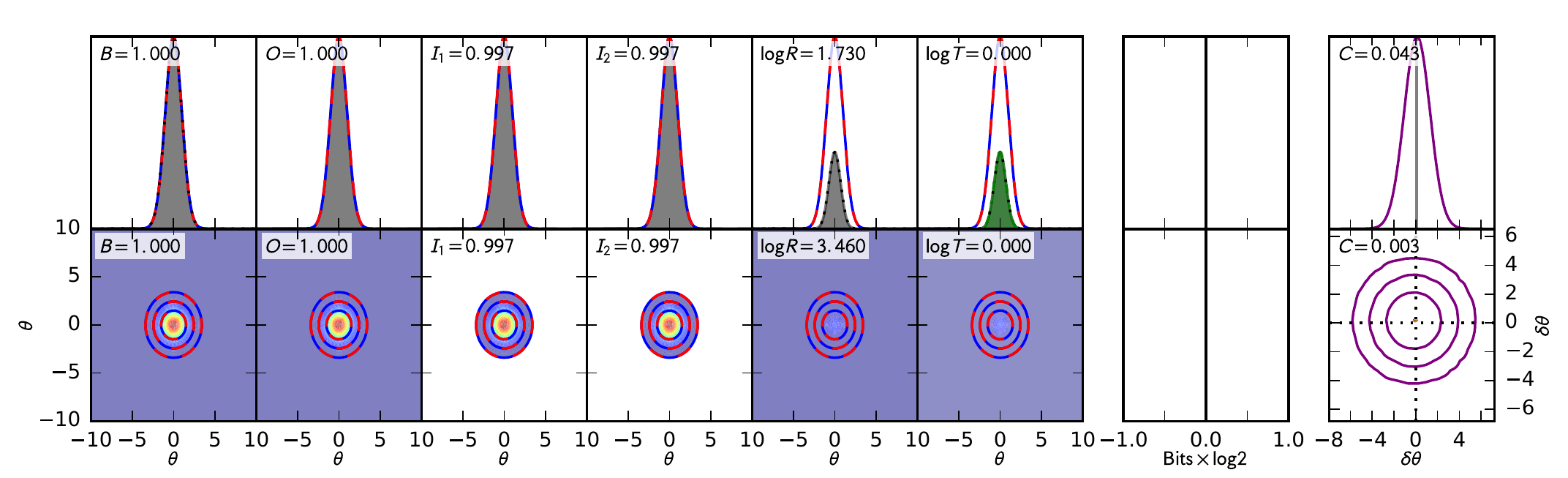}
                     \caption[Comparison of identical distributions]{Comparison of identical distributions (I)\label{fig:I}}
                     \end{figure*}

\begin{figure*}
                     \centering
                     \includegraphics[width=\textwidth]{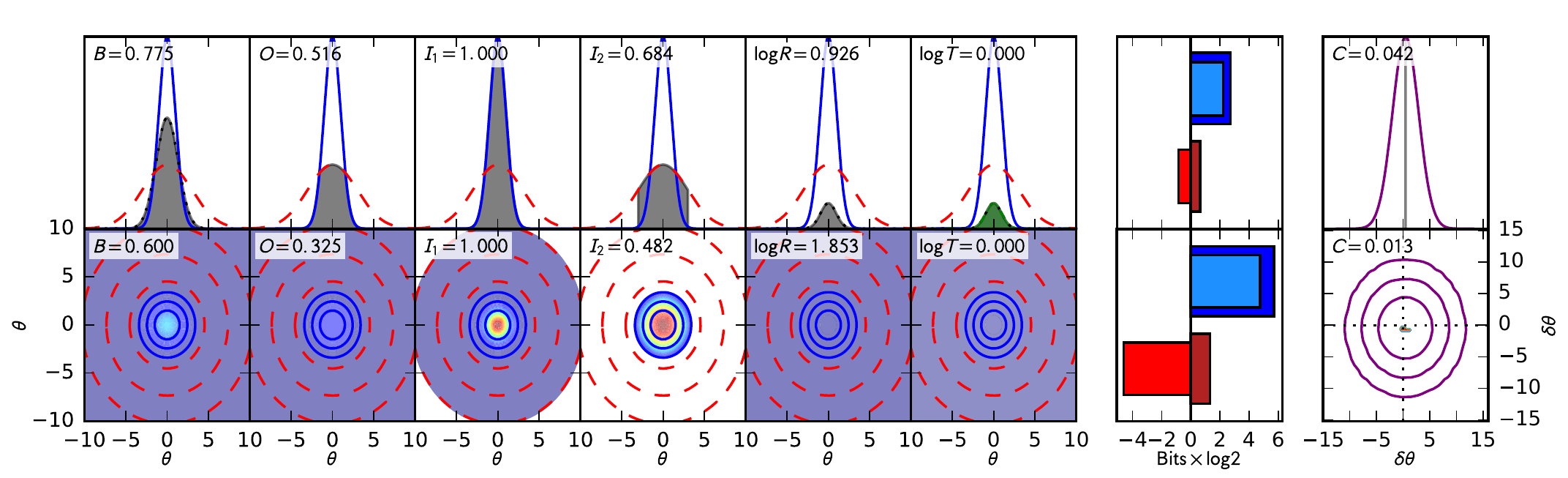}
                     \caption[Comparison of one distribution broader than the other but with the same means]{Comparison of one distribution broader than the other but with the same means (II) \label{fig:II}}
                 \end{figure*}

\begin{figure*}
                     \centering
                     \includegraphics[width=\textwidth]{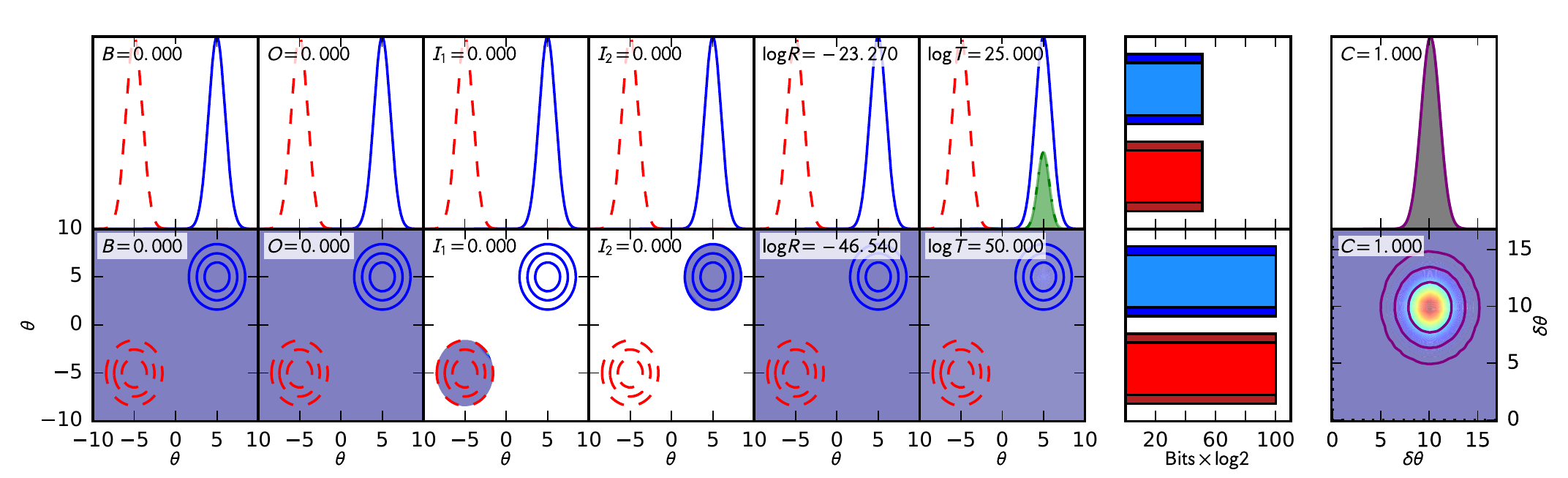}
                     \caption[Comparison of discordant probability distributions]{Comparison of discordant probability distributions (III)\label{fig:III}}
                     \end{figure*}

\begin{figure*}
                     \includegraphics[width=\textwidth]{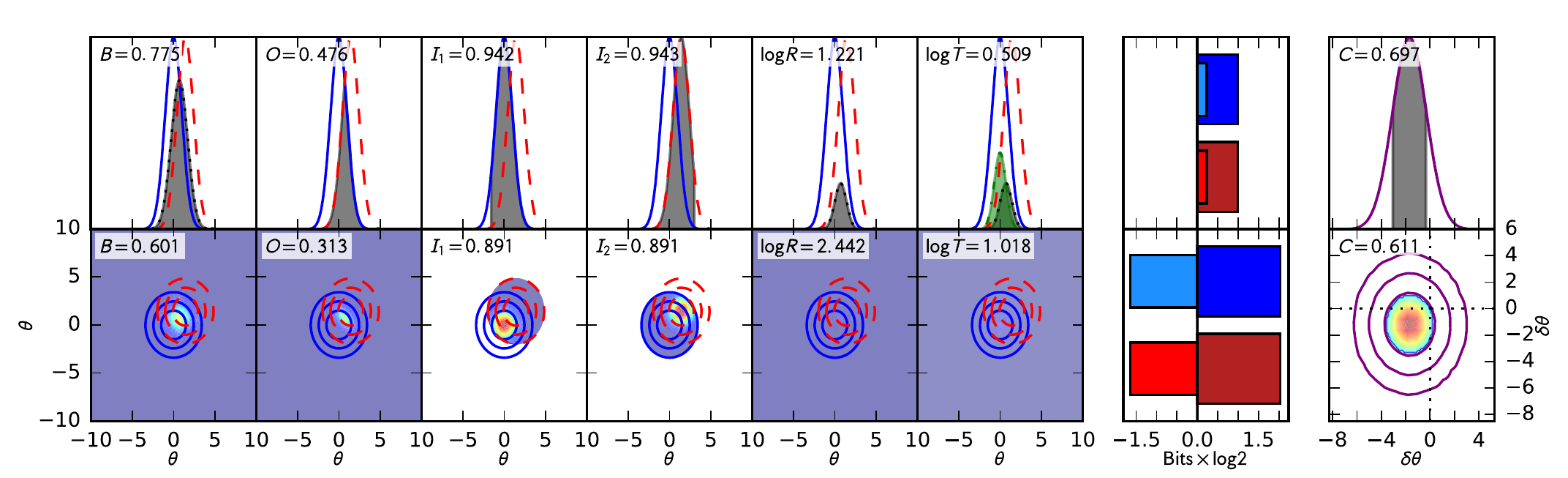}
                     \caption[Comparison of slightly shifted distributions]{Comparison of slightly shifted distributions (IV)\label{fig:IV}}
                \end{figure*}

\begin{figure*}
                     \centering
                     \includegraphics[width=\textwidth]{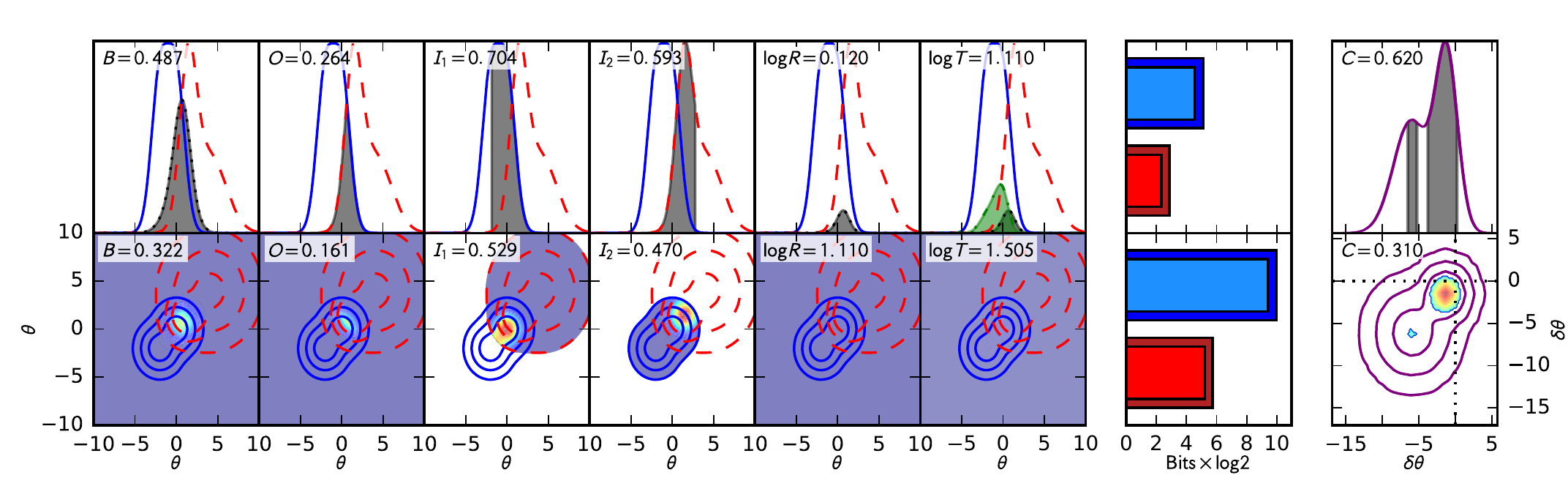}
                     \caption[Comparison of unusually shaped distributions]{Comparison of unusually shaped distributions (V)\label{fig:V}}
                 \end{figure*}

                 Each figure in this section shows the comparison of two probability distributions for each method discussed in the previous section.
                 The top row of figures~\ref{fig:I}, \ref{fig:II}, \ref{fig:III}, \ref{fig:IV} and~\ref{fig:V} show the comparisons of the distributions in rows~I, II, III, IV and V in table~\ref{t:1Ddistributions} respectively.
                 Likewise, the bottom row of each figure shows the comparison between the distributions in rows~I --~V in table~\ref{t:2Ddistributions}.
                 The columns show the Bhattacharyya distance (\emph{1}), the overlap coefficient (\emph{2}), the integral of $P_1$ between the limits containing 99.7\% of $P_2$ and the integral of $P_2$ between the limits containing 99.7\% of $P_1$ (\emph{4}), the quantification of Bayesian evidence (\emph{6}), the shifted probability distribution (\emph{7}), ``surprise'' (\emph{5}) and the difference vector (\emph{3}) from left to right.
                 For the first six columns the solid, blue and dashed, red lines indicate the distributions $P_1$ and $P_2$ respectively.
                 In the top rows, the shaded grey area (bounded by a dotted, black line) shows the integrated quantity used to give the comparison measure.
                 In the bottom rows, the integrated quantities are shaded with blue being close to zero, turning red for ${\rm Max}[P_1,P_2]$.
                 In the top row of the sixth column the green shaded area (bounded by a dot-dashed, green line) indicates the integrated shifted quantity $P_1P^{~\rm shifted}_2$, whilst the grey shaded area (bounded by a dotted, black line) marks the integrated non-shifted quantity $P_1P_2$, the ratio of which gives the measure.
                 The seventh column shows the amount of relative entropy in the wider, darker bars and the amount of ``surprise'' in the slimmer, lighter bars.
                 The upper, blue bars indicate the relative entropy and ``surprise'' when $P_2$ is used to update $P_1$ and the lower, red bars show the relative entropy and ``surprise'' when $P_1$ updates $P_2$.
                 The final column shows the probability distribution of the difference vector with a solid purple line.
                 The grey shaded area in the top row is the integrated quantity giving the measure.
                 The integration bounds are the values of the probability distribution greater than its value at $\dtheta=0$.
    \end{appendices}

\end{document}